\documentclass[aip,pof,reprint,amsmath,amssymb,longbibliography]{revtex4-2}

\usepackage{float}
\usepackage{amsfonts}
\usepackage{bm}
\usepackage{graphicx}
\usepackage{booktabs}
\usepackage{algorithm}
\usepackage{algorithmic}
\usepackage{xcolor}
\usepackage[colorlinks=true,linkcolor=blue,citecolor=blue,urlcolor=blue]{hyperref}

\begin{document}

\title{Physics-informed wavelet--Fourier representation for multiscale fluid dynamics}

\author{Chao Wang}
\affiliation{College of Shipbuilding Engineering, Harbin Engineering University, Harbin 150001, China}
\affiliation{Qingdao Innovation and Development Base, Harbin Engineering University, Qingdao 266000, China}
\affiliation{National Key Laboratory of Hydrodynamics, Harbin Engineering University, Harbin 150001, China}

\author{Shilong Li}
\affiliation{College of Shipbuilding Engineering, Harbin Engineering University, Harbin 150001, China}
\affiliation{Qingdao Innovation and Development Base, Harbin Engineering University, Qingdao 266000, China}

\author{Yunpeng Wang}
\affiliation{Department of Mechanics and Aerospace Engineering, Southern University of Science and Technology, Shenzhen 518055, China}

\author{Tianbai Xiao}
\affiliation{Centre for Interdisciplinary Research in Fluids, Institute of Mechanics, Chinese Academy of Sciences, Beijing 100190, China}

\author{Zelong Yuan}
\email{yuanzelong@hrbeu.edu.cn}
\affiliation{College of Shipbuilding Engineering, Harbin Engineering University, Harbin 150001, China}
\affiliation{Qingdao Innovation and Development Base, Harbin Engineering University, Qingdao 266000, China}
\affiliation{Nanhai Institute of Harbin Engineering University, Harbin Engineering University, Sanya 572024, China}
\affiliation{National Key Laboratory of Hydrodynamics, Harbin Engineering University, Harbin 150001, China}

\author{Chenyue Xie}
\email{cyxie@ustc.edu.cn}
\affiliation{Department of Modern Mechanics, University of Science and Technology of China, Hefei 230026, China}

\author{Chunyu Guo}
\affiliation{College of Shipbuilding Engineering, Harbin Engineering University, Harbin 150001, China}
\affiliation{Qingdao Innovation and Development Base, Harbin Engineering University, Qingdao 266000, China}
\affiliation{Nanhai Institute of Harbin Engineering University, Harbin Engineering University, Sanya 572024, China}
\affiliation{National Key Laboratory of Hydrodynamics, Harbin Engineering University, Harbin 150001, China}

\date{\today}

\begin{abstract}
Multiscale fluid flows often contain localized flow structures, such as viscous shock layers, wet--dry fronts, steady viscous wakes, decaying vortical structures, and vortex-shedding patterns, whose accurate prediction requires the simultaneous preservation of global conservation trends and small-scale gradients. This study examines these flow-physics requirements through a physics-informed wavelet--Fourier (PIWF) representation for multiscale fluid dynamics. Instead of relying on a single monolithic neural approximator, the formulation separates two complementary components of the flow field within a physics-informed neural representation: long-range coherent modes through a Fourier-basis branch and localized steep-gradient or vortical features through a compactly supported wavelet branch. The outputs are fused with a residual multilayer perceptron using channel attention, and the governing equations, initial conditions, and boundary conditions are imposed directly through the physics-informed loss. The model is assessed on five canonical fluid-dynamics problems: Burgers' equation, the shallow water equations, Kovasznay flow, Taylor--Green vortex flow, and two-dimensional cylinder wake flow. The results show that PIWF improves the resolution of shock-like gradients, wet--dry interfaces, steady wake fields, decaying vortical structures, vorticity extrema, and broadband wake spectra relative to standard physics-informed neural networks and physics-informed Kolmogorov--Arnold networks. These findings indicate that a wavelet--Fourier physics-informed representation can provide a useful route for analyzing multiscale flow phenomena when high-fidelity interior reference data are limited or unavailable.
\end{abstract}


\maketitle

\section{\label{sec:level1}Introduction}
Multiscale fluid flows are organized by interacting structures across widely separated length and time scales, including shock-like gradients, wet--dry fronts, coherent vortices, and broadband wake motions. Partial differential equations (PDEs) provide the fundamental mathematical framework for describing a broad class of physical phenomena and engineering applications \cite{strauss2007partial}, including fluid dynamics \cite{pearson1965computational}, wave propagation \cite{griffiths2010traveling}, reactive transport \cite{christofides2002nonlinear,steefel1996approaches}, and multiphysics coupling \cite{errera2011multi}. In the absence of analytical solutions for most practical problems, classical numerical methods such as finite element methods \cite{kang1996finite}, finite difference methods \cite{ozicsik2017finite}, and finite volume methods \cite{moukalled2015finite} have been widely employed and have achieved remarkable success \cite{moin1998direct}. However, these classical numerical methods face fundamental limitations in real-world PDE problems \cite{liu2024multi,bauerheim2026routes}. Multiscale dynamics demand extremely fine resolutions to capture sharp gradients, discontinuities, and singularities, leading to the curse of dimensionality and prohibitively high computational cost \cite{jebahi2016multiscale}. Complex geometries require manual and time-consuming mesh generation, while strongly coupled nonlinear processes introduce stiffness, stability issues, and conservation difficulties \cite{zhao2026review}. Consequently, the development of robust, flexible, and high-fidelity numerical solvers remains critical for advancing scientific discovery and engineering applications \cite{guruswamy2002review}. For fluid-dynamics applications, this need is closely tied to the faithful representation of the flow structures that control transport, separation, front propagation, and vortex dynamics.

In this context, physics-informed and data-augmented approaches have emerged as promising alternatives that effectively balance predictive accuracy with physical consistency. These methods facilitate flexible exploration of parameter spaces, inverse problems, and multi-fidelity modeling \cite{raissi2019physics}. In particular, physics-informed neural networks (PINNs) have gained considerable attention by embedding the governing PDEs directly into the training loss via automatic differentiation. This approach yields mesh-free and differentiable solvers that can incorporate sparse or noisy data while strictly enforcing physical laws \cite{zhao2025lesnets,wang2017physics,jin2021nsfnets,arzani2021uncovering}. For fluid-dynamics problems, recent physics-informed studies have emphasized that predictive accuracy should be assessed together with the consistency of the recovered multiscale flow structures, rather than only through aggregate error metrics \cite{zhao2024comprehensivepof,zhou2024advancingpinn,wu2025khpinn}. PINNs have been successfully applied to a wide range of forward and inverse problems \cite{lu2021physics}. In parameter estimation tasks, they enable the simultaneous inference of the solution to the governing PDEs and the identification of unknown coefficients or parameters directly from sparse and noisy observations \cite{raissi2019physics}. In topology optimization, PINNs facilitate mesh-free, gradient-based design of material distributions by embedding physical constraints into the loss function \cite{jeong2023physics}. For multiphysics couplings, they provide a unified framework for handling strongly coupled systems (like microfluidic problems) without requiring separate solvers or explicit interface conditions \cite{sun2024physics}. These applications often provide substantial practical advantages over conventional mesh-based solvers in terms of flexibility, ease of implementation, and handling of complex geometries \cite{luo2025physics,wang2024recent,hao2022physics}. 

However, standard PINN architectures struggle to simultaneously capture global structures and small-scale local features, especially in problems involving sharp interfaces, vortical structures, or highly intermittent dynamics \cite{lu2026r,zou2025uncertainty,kang2026asymptotic,zhao2024comprehensive}. A key challenge is that neural networks tend to capture dominant global modes more readily than localized phenomena, reflecting their spectral biases in fixed-activation networks, which might lead to slow convergence and suboptimal accuracy in multiscale regimes \cite{ren2023physr}. To circumvent these limitations, various strategies have been proposed. These include adaptive sampling techniques that dynamically refine collocation points near discontinuities, the explicit incorporation of Rankine--Hugoniot jump conditions for conservation laws, domain decomposition methods, and the use of multiscale activation functions or positional encodings. While these approaches can alleviate spectral bias to some extent, each comes with its own trade-offs, such as increased implementation complexity, problem-specific tuning, or additional computational overhead.
To circumvent these limitations, recent research has shifted toward architectures that depart from the traditional fixed-node paradigm to more effectively extract multiscale information without prohibitive computational overhead.
Another promising direction is the Kolmogorov--Arnold network (KAN) introduced by Liu et al. \cite{liu2024kan}, which represents a fundamental shift in neural-network topology by replacing fixed activation functions on nodes with learnable, spline-based functions on the network edges. This structural innovation allows for more flexible nonlinear mapping and has demonstrated superior Pareto frontiers in both data-fitting and PDE benchmarks. Building upon this framework to further enhance numerical stability and spectral accuracy, Guo et al. \cite{guo2025physics} integrated orthogonal Chebyshev polynomials into the KAN architecture. Bozorgasl et al. proposed the wavelet Kolmogorov-Arnold networks \cite{bozorgasl2405wav}, which leverages wavelet transforms to efficiently capture both high- and low-frequency components of data, significantly improving predictive accuracy, training efficiency, and inference robustness compared to KAN and MLPs. Aghaei et al. proposed the rational Kolmogorov-Arnold networks \cite{aghaei2024rkan}, which introduces rational basis functions based on Pade approximation and rational Jacobi functions, demonstrating superior performance in function approximation across deep learning and physics-informed tasks.

Although KAN-based architectures enhance the network's ability to capture multiscale features, the underlying B-spline basis functions are inherently ill-suited for approximating discontinuities and abrupt changes, limiting their effectiveness in representing extremely localized multiscale phenomena. In such cases, wavelet analysis has been recognized as a powerful tool for representing time series and signals \cite{zhou2020modeling,chang2000adaptive}, which excels at capturing localized features and nonlinearities \cite{cao1995predicting,fang2006wavelets}. In the context of PDE solving, Harnish et al. \cite{harnish2023adaptive} developed an adaptive wavelet method for coupled nonlinear partial differential equations. By exploiting the multi-resolution properties of wavelet bases, their approach dynamically adapts the computational grid, maintains high accuracy, and achieves substantial data compression, with successful applications in dynamic damage modeling. However, classical wavelet-based methods are typically restricted to low-dimensional problems, as the explicit construction of wavelet bases becomes computationally prohibitive in high-dimensional spaces \cite{liu2019wavelet,yang2024high}. 

In contrast, neural networks can approximate arbitrary nonlinear functions without requiring strong prior assumptions about the underlying dynamics \cite{frasconi1995recurrent}. Nevertheless, traditional neural architectures may encounter optimization difficulties, including vanishing or poorly scaled gradients, when representing multiscale solution fields with sharp transitions. As a result, their training can become slow, initialization-sensitive, and prone to inaccurate reconstruction of localized high-frequency or high-gradient features \cite{szandala2020review}. These limitations have motivated the integration of wavelet techniques into neural architectures. Alexandridis et al. \cite{alexandridis2013wavelet} established a comprehensive statistical framework for wavelet networks, addressing network structure, training algorithms, initialization, variable selection, and model validation. Navaneeth et al. \cite{navaneeth2024physics} proposed the physics-informed wavelet neural operator for learning solution operators for parametric PDEs without requiring high-fidelity training data. Gu et al. \cite{gu2026physics} introduced the physics-informed multiscale attention wavelet neural operator as a surrogate model for linear parabolized stability equations, demonstrating strong performance on canonical instability modes in hypersonic boundary layers.

Despite these promising advances, each approach carries its own limitations when applied to physics-informed modeling of multiscale phenomena. KANs, despite their improved PDE accuracy, rely on B-spline basis functions that are naturally suited for approximating smooth functions. They become less effective in resolving sharp local discontinuities or abrupt transitions, as spline representations tend to smooth out high-frequency oscillations. This architecture makes it difficult to balance the extraction of global structures and localized features simultaneously, especially when confronted with shocks or highly intermittent dynamics. Existing approaches still struggle to simultaneously capture global large-scale structures and localized high-frequency features within a single physics-informed framework, especially when high-fidelity reference data are scarce or unavailable. This persistent trade-off between global consistency and local detail, inherited from the spectral bias of standard PINNs and the dimensionality limitations of classical wavelets, limits their robustness for practical multiscale PDE problems.
To address the aforementioned limitations, we propose the physics-informed wavelet--Fourier (PIWF) model, a scale-aware representation for multiscale fluid dynamics that synergistically integrates multiscale wavelet representations with physics-informed learning. The core innovation lies in effectively bridging the persistent trade-off between global coherence and local fidelity that constrains conventional PINNs. Specifically, PIWF incorporates a dedicated wavelet layer that embeds wavelet series expansions directly into the network, enabling accurate capture of localized high-frequency features such as sharp interfaces and coherent vortical structures. This is complemented by two parallel branches: a Fourier-basis layer \cite{wang2025physics} for efficient extraction of global features and a conventional multi-layer perceptron (MLP) for residual compensation and information integration. Channel attention \cite{qin2021fcanet} is then employed to adaptively fuse the outputs from these branches, enabling a data-driven balance between global consistency and local fidelity. By embedding the governing physical laws into the loss function, PIWF can be trained with little or no high-fidelity interior reference data. The emphasis is therefore on resolving physically meaningful multiscale flow structures.

The remainder of this paper is organized as follows: Section \ref{sec:level1/2} presents the proposed PIWF architecture, including the formulation of the wavelet layer, the integration of the Fourier-basis layer and MLP branches, the channel attention fusion mechanism, and the incorporation of physics-informed constraints. Section \ref{sec:level1/3} assesses PIWF through numerical experiments on several canonical fluid dynamics problems, including Burgers' equation, the shallow-water equations, Kovasznay flow, Taylor--Green vortex flow, and the two-dimensional cylinder wake flow. Finally, conclusions and outlooks are discussed in Section \ref{sec:level1/6}.

\section{\label{sec:level1/2}Methodology}

The PIWF formulation is constructed to reflect a common structure of multiscale fluid solutions: large-scale coherent motion coexists with localized fronts, shear layers, shocks, or vortex cores. A discrete wavelet expansion provides localized multiresolution support for the latter features, whereas Fourier bases efficiently represent spatially extended and smoothly varying modes. We therefore reformulate the conventional fully connected representation into three parallel components: a wavelet layer for local high-gradient content, a Fourier-basis layer \cite{wang2025physics} for global coherent content, and a residual MLP branch for nonlinear compensation. Their outputs are fused through channel attention so that the relative contribution of each scale-aware representation can vary during training. The resulting physics-informed wavelet--Fourier model is designed for flow fields governed by conservation laws while retaining interpretable links between the numerical representation and the physical structures being resolved.

\subsection{\label{sec:level3}Physics-informed neural network (PINN)}
The core idea of physics-informed neural networks is to incorporate physical governing equations into neural network models. This allows PINNs to overcome key limitations of purely data-driven methods by enforcing physical consistency in the model outputs. Consider a general form of nonlinear PDE,
\begin{align}
	\frac{\partial \bm{u}}{\partial t} & = R(\bm{u}; \bm{x},t),  &\text{in } \Omega \times [T_0, T], \label{eq11} \\ 
	\mathcal{B}(\bm{u}) & = \bm{u}_{g}(\bm{x},t),  &\text{on } \partial \Omega \times [T_0, T], \label{eq12} \\ 
	\mathcal{I}(\bm{u}) & = \bm{u}_{a}(\bm{x},t),  &\text{in } \Omega \times \{ T_0 \}, \label{eq13} 
\end{align}
where $\Omega$ is the spatial domain, and $[T_0,T]$ is the time interval, with $T_0$ and $T$ denoting the initial and terminal time, respectively. The operators $\mathcal{B}(u)$ and $\mathcal{I}(u)$ represent the boundary and initial conditions, respectively. The parameters of the neural network are optimized by minimizing the composite loss function,
\begin{equation}
	\begin{split}
		\mathcal{L} ( \theta ) =& \underbrace { \frac { \lambda_1} { N_{\rm{ic}} } \sum _ { i = 1 } ^ { N_{\rm{ic}} } || \mathcal{I}[\bm{u}_{ \theta } ( \bm{x} ^ { i } ,T_0 )] - \bm{u}_{a} (\bm{x} ^ { i } ,T_0) || ^ { 2 } } _ {\mathcal L_{\rm{ic}} ( \theta ) } +
		\underbrace { \frac { \lambda_2} { N_{\rm{bc}} } \sum _ { i = 1 } ^ { N_{\rm{bc}} } || 	\mathcal{B}[\bm{u}_{ \theta } ( t^i , \bm{x}_{\rm{bc}} ^ { i } )] - \bm{u}_{g} (\bm{x}_{\rm{bc}} ^ { i },t^i ) || ^ { 2 } } _ {\mathcal L_{\rm{bc}} ( \theta ) } \\+& 
		\underbrace { \frac { \lambda_3 } { N_{\rm{pde}} } \sum _ { i = 1 } ^ { N_{\rm{pde}} } ||\mathcal L_{\rm pde} ( \bm{u} _ { \theta }  ,\bm{x}^{i}_r,t^{i}_r  ) || ^ { 2 } } _ {\mathcal L_{\rm{pde}} ( \theta ) },
		\label{16}
	\end{split}
\end{equation}
where $\bm{x}^i \in \Omega$, $\bm{x}^{i}_{\rm{bc}} \in \partial \Omega$, $t^i \in [T_0,T]$, and $\bm{u}_{\theta}$ represents the outputs of neural networks. $\bm{u}_g$ and $\bm{u}_a$ denote the boundary conditions and initial conditions, respectively. $N_{\rm{ic}}$, $N_{\rm{bc}}$ and $N_{\rm{pde}}$ are the numbers of sampling points corresponding to initial conditions, boundary conditions, and collocation points, respectively. $\lambda_1, \lambda_2$ and $\lambda_3$ are the correspondingly given weights of each loss. By substituting the input pair of collocation points for PDE $(\bm{x}_ {\rm{pde}},t_{\rm {pde}})$ and model prediction $\bm{u}_\theta(\bm{x}_ {\rm{pde}},t_{\rm {pde}})$ into Eq. \ref{eq11}, the equation residual can be obtained through automatic differentiation, namely
\begin{equation}
	\mathcal{L}_{\rm pde} ( \bm{u} _ { \theta }  ;\bm{x}_{ \rm{pde} },t_{ \rm{pde} } ) = \frac { \partial \bm{u} _ { \theta }}  { \partial t } ( \bm{x} _ { \rm{pde} } ,t _ { \rm{pde} } ) - R ( \bm{u} _ { \theta } ;\bm{x} _ { \rm{pde} },t _ { \rm{pde}}  ).
	\label{eq15}
\end{equation}
The standard PINN employs a fully connected multi-layer perceptron (MLP) as the backbone to approximate the solution $  \bm{u}(\bm{x},t)  $. 
In this work, the MLP consists of an input layer, $  L  $ hidden layers, and an output layer. Let $  \bm{a} = (\bm{x},t)  $ denote the input coordinate vector. The neural network is defined recursively by
\begin{equation}
	\begin{aligned}
		&\bm{y}^{(0)} = \bm{a}, \\
		&\bm{y}^{(l)} = \sigma\left(\mathbf{W}^{(l)}\bm{y}^{(l-1)} + \bm{b}^{(l)}\right), \quad l=1,2,\dots,L-1, \\
		&\bm{u}_{\theta}(\bm{x}, t) = \mathbf{W}^{(L)}\bm{y}^{(L-1)} + \bm{b}^{(L)},
	\end{aligned}
\end{equation}
where $  \mathbf{W}^{(l)}\in\mathbb{R}^{N_l\times N_{l-1}}  $ and $  \bm{b}^{(l)}\in\mathbb{R}^{N_l}  $ are the weight matrix and bias vector of the $  l  $-th layer, $  N_l  $ is the number of neurons in the $  l  $-th layer, and $  \sigma(\cdot) $ is the activation function. The trainable parameters $  \theta=\{\mathbf{W}^{(l)},\bm{b}^{(l)}\}_{l=1}^{L}  $ are optimized by minimizing the composite loss in Eq. \ref{16} using the Adam optimizer \cite{Adam}, with gradients evaluated by automatic differentiation.

\subsection{\label{sec:level-wavelet}Wavelet layers}
Conventional fully-connected layers in multi-layer perceptrons are expressed as $\bm{y} = \sigma ( \mathbf{W} \bm{x} + \bm{b} )$, where $\sigma$ denotes a nonlinear activation function. Although effective for general function approximation, this structure treats the input as a flattened vector, assigning equal importance to all data points and largely neglecting underlying spatial, temporal, or topological structures. As a result, it often fails to effectively capture localized features with sharp gradients.

To address this limitation, we introduce a dedicated wavelet layer that emulates the expansion of a wavelet series within the neural-network mapping. In the present fluid-dynamics context, the purpose of this layer is to provide localized basis support for abrupt spatial changes, including shock-like gradients, wet--dry interfaces, and compact vortical features. The core idea is to utilize a neural network to emulate the wavelet series in Eq. \ref{wavelet} of Appendix~\ref{sec:level3wavelt}, thereby achieving function approximation at the layer level. Specifically, given input-output pairs, there exists a unique mapping $f(\bm{x}):\mathbb{R}^{d_x} \mapsto \mathbb{R}^{d_y}$, where $d_x$ and $d_y$ denote the dimensions of the input and output spaces, respectively. The role of the neural network is to act as a surrogate model for this unique mapping, such that $\bm{y}_i = f(\bm{x}_i) \approx f_\theta(\bm{x}_i)$, where $ f_\theta(\bm{x}_i)$ is the neural network function with parameters $\theta$ to be optimized.

As detailed in Appendix~\ref{sec:level3wavelt}, the wavelet-series representation is employed to approximate the target input-output mapping 
$\bm{x}_i \mapsto \bm{y}_i$. This formulation enables localized multi-resolution approximation and provides the basis for the proposed wavelet layer. Our objective is to construct the neural network layer to  emulate the mathematical formulation of this wavelet series, namely
\begin{equation}
	f_\theta(\bm{x}_i)  = \sum _ { k } c _ { J , k } \phi _ { J , k } ( \bm{x}_i ) + \sum _ { j = J } ^ { \infty } \sum _ { k } d _ { j , k } \psi _ { j , k } ( \bm{x}_i ).
\end{equation}
To alleviate computational redundancy and enhance efficiency, the wavelet series is further simplified as follows.

\romannumeral 1. Regarding the scaling function, the decomposition level $J $ is chosen such that the dilated width of the scaling function $\phi ( 2 ^ { -j } \bm{x} - k )$ is greater than or equal to the entire computational domain. Assuming the mother scaling function is compactly supported on the interval $[0,L]$, it follows from Eq. \ref{waveletmom} in Appendix~\ref{sec:level3wavelt} that the support of the dilated mother scaling function is given by
\begin{equation}
	0 \leq 2 ^ { -j }  \cdot \bm{x} - k < L.
\end{equation}
Solving this inequality for $\bm{x}$ gives
\begin{equation}
	k \cdot 2^j \leq  \bm{x}  < (k + L) \cdot 2^j.
\end{equation}
Therefore, the support interval after dilation is $[k \cdot 2^j, (k + L) \cdot 2^j]$, where the width of the dilated wavelet is
\begin{equation}
	(k + L) \cdot 2^j - k \cdot 2^j = L \cdot 2^j.
\end{equation}
Assuming the width of the computational domain is $T$, it requires that $T \leq L \cdot 2^j$. This configuration ensures that the computation of the scaling function involves only a single term corresponding to $k=0$. Furthermore, after dilation at the coarsest scale $J$, the wavelet spans the entire computational domain and therefore cannot be further translated. This design forces the network layer to concentrate its resources on learning the global information of the field. Without this simplification, the network would suffer not only from increased computational redundancy, but also from a greater tendency to overfit local details.

\romannumeral 2.  For the detail space, we restrict $k\in (-2,2)$ and arrange the indices $j$ in ascending order, applying a truncation. This arrangement enables the wavelet basis functions to be learned sequentially, from coarse to fine details. An excessive number of detail space wavelets would cause the neural network to over-emphasize the learning of local details at the expense of global information. Concurrently, this requires the coarsest scale level $J$ to be sufficiently large, otherwise, for specific problems, the fixed range of $k$ and an insufficiently small $j$ would result in wavelets that can only be dilated and translated within the central region of the computational domain, failing to cover its boundaries. Unless otherwise specified, we set $J=3$ throughout this paper. With these simplifications, the resulting streamlined network layer can be written as follows
\begin{equation}
	f_\theta(\bm{x}_i)  =  c _ { J , 0 } \phi _ { J , 0 } ( \bm{x}_i ) + \sum _ { j = 0 } ^ { J } \sum _ { k } d _ { j , k } \psi _ { j , k } ( \bm{x}_i ).
\end{equation}
Due to the applied truncation with a cutoff number of $N$, we reformulate the network layer into a matrix form, yielding
\begin{equation}
	\begin{aligned}
		f_\theta(\bm{x})  &=  {{\mathbf C} _ { J , 0 }} \phi _ { J , 0 } (\bm{x} ) + \sum _ { n = 0 } ^ { N-1 }  {{\mathbf D} _ { n }} \psi _ { n } ( \bm{x}) \\
		&= [{{\mathbf C} _ { J , 0 }},\mathbf{D}_0,\mathbf{D}_1,...,\mathbf{D}_{N-1}] [\phi _ { J , 0 } (\bm{x} ), \psi _ { 0 } ( \bm{x}),\psi _ { 1 } ( \bm{x}),...,\psi _ { N-1 } ( \bm{x})]^T\\
		&= \mathbf{W}_1[\phi _ { J , 0 } (\bm{x} ), \psi _ { 0 } ( \bm{x}),\psi _ { 1 } ( \bm{x}),...,\psi _ { N-1 } ( \bm{x})]^T,
	\end{aligned}
\end{equation}
where $\mathbf{C}_{ J , 0 }\in \mathbb{R} ^ { d_{\theta} \times d_x  }$ ,$\mathbf{D}_n\in \mathbb{R} ^ { d_{\theta} \times  d_x  }$,$\mathbf{W}_1\in \mathbb{R} ^ {   d_{\theta} \times(N+1)\cdot d_x  }$, $[...]^T$ and $[...]$ denote the concatenations along the first and second dimensions, respectively, $d_x$ is the dimension of input $\bm{x}$, and $d_{\theta}$ represents the number of nodes in the network layer.

To compensate for approximation errors introduced by basis functions, we combine the wavelet branch with a parallel fully-connected residual branch. The complete output of the wavelet layer is then
\begin{equation}
	{\bm y}_w(\bm{x}) \triangleq 
	\begin{bmatrix}
		\mathbf{W}_1 \Bigl[ \phi_{J,0}(\bm{x}), \, 
		\psi_0(\bm{x}), \, \psi_1(\bm{x}), \, \dots, \, \psi_{N-1}(\bm{x}) \Bigr] \\
		\sigma \bigl( \mathbf{B} + \mathbf{W}_2 \bm{x} \bigr)
	\end{bmatrix},
\end{equation}
where $\mathbf{W}_1\in\mathbb{R}^{d_{q}\times(N+1)\cdot d_x}$ and $\mathbf{W}_2\in\mathbb{R}^{   (d_{\theta}-d_q)\times(N+1)\cdot d_x}$. The first block corresponds to the wavelet contribution and the second to the residual MLP branch.
The computation of the wavelet basis functions is implemented over the nodes of the network layer. Therefore, for each layer, a hyperparameter $r_w$ is introduced to define the ratio between the number of nodes allocated for computing the wavelet basis functions and the number of nodes in the standard fully-connected component, namely $r_w \cdot d_{\theta}= d_q$.

\begin{figure}[htbp] 
	\centering
	\includegraphics[width=1\textwidth]{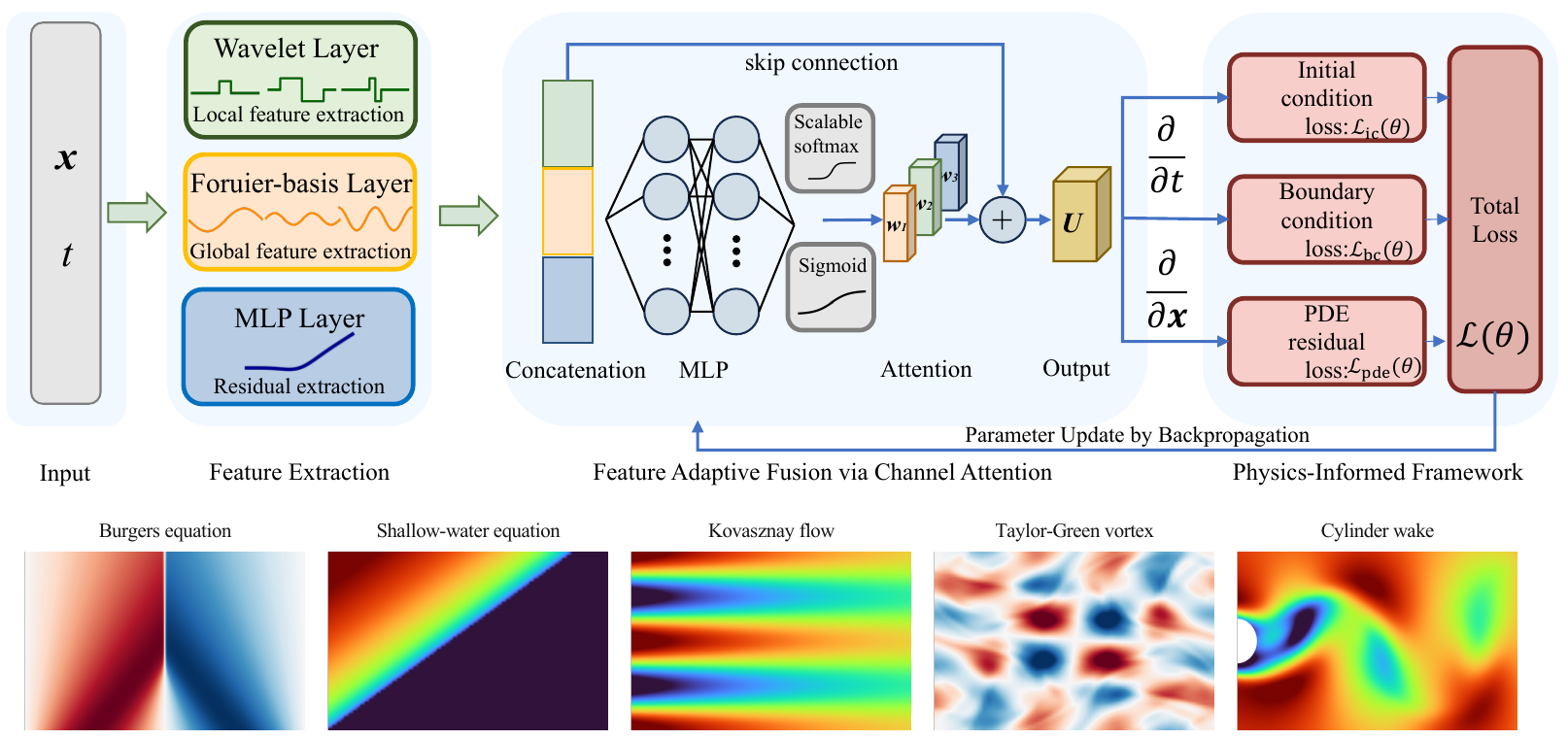}
	\caption{The physics-informed wavelet--Fourier (PIWF) representation for multiscale fluid dynamics.}
	\label{architecture}
\end{figure}

In this work, the Haar wavelet function is adopted as the mother wavelet because its compact support, orthogonality, symmetry, and computational efficiency make it well suited for steep and localized flow features. In contrast to smooth global bases, Haar functions can represent sudden changes without spreading the correction over the entire domain, which is desirable for shocks, wet--dry fronts, and localized extrema in derivative-based flow diagnostics. The scaling and wavelet functions of the Haar wavelet are defined by
\begin{equation}
	\phi(t) = 
	\begin{cases} 
		1, & 0 \le t < 1, \\
		0, & \text{otherwise},
	\end{cases}
\end{equation}
\begin{equation}
	\psi(t) = 
	\begin{cases} 
		1, & 0 \le t < \frac{1}{2}, \\
		-1, & \frac{1}{2} \le t < 1, \\
		0, & \text{otherwise}.
	\end{cases}
\end{equation}
The Haar wavelet function enables precise localization of singularities while maintaining sparsity and avoiding phase distortion, making it particularly suitable for capturing sharp interfaces and steep-gradient regions in fluid dynamics problems \cite{guo2022review}.

\subsection{\label{sec:level-FBNN-a}Fourier-basis layer}
Given input-output pairs $\{ \bm{x}_i, \bm{y}_i \}$, a neural network aims to learn a mapping $f_\theta(\bm{x}):\mathbb{R}^{d_x} \mapsto \mathbb{R}^{d_y}$, such that $\bm{y}_i = f(\bm{x}_i) \approx f_\theta(\bm{x}_i)$. 
In the Fourier-basis layer, this mapping is approximated by a truncated Fourier series, namely
\begin{equation}
	f(\bm x_i) \approx  A _ { 0 } + \sum _ { n = 1 } ^ {N} \left( A _ { n } \cos \left( \frac { 2 \pi n \bm x_i } { L_\Omega } \right) + B _ { n } \sin \left( \frac { 2 \pi n \bm x_i } { L_\Omega } \right) \right) \approx f_\theta(\bm x_i),
	\label{eq:fit}
\end{equation}
where $N$ is the order of the Fourier series. $L_\Omega$ is the domain length. The Fourier coefficients ($A_0$, $ A_n$, and $B_n$) are given by
\begin{equation}
	A_n = \frac{1}{L_\Omega} \int_0^{L_\Omega} f(\bm{\bm{x}}) \cos\left( \frac{2\pi n \bm{x}}{L_\Omega} \right) dx, \quad {\rm{and}} \quad
	B_n = \frac{1}{L_\Omega} \int_0^{L_\Omega} f(\bm{x}) \sin\left( \frac{2\pi n \bm{x}}{L_\Omega} \right) dx.
	\label{eq:fourier_final}
\end{equation}
To implement Eq. \ref{eq:fit} within a neural network layer, the coefficients $a_n^l$ and $b_n^l$  respectively represent the Fourier coefficients of the $l$-th ($l \in [0, L]$) layer of Fourier basis neural network and a weight matrix $\bm{W}^l=[w_1^l,w_2^l,w_3^l,\dots,w_N^l]^T$ represents the angular frequencies ${2 \pi n }/{ L_\Omega } $ in the Fourier series
\begin{align}
	f^{l}_\theta(\bm x_i) &\approx A_0 + \sum_{n=1}^{N} \left( a_n^l \cos \left( w_n^l \bm x_i \right) + b_n^l \sin \left( w_n^l \bm x_i \right) \right) \notag \\
	&= A_0 + [a_1^l, a_2^l, \dots, a_N^l] \cos([w_1^l, w_2^l, \dots, w_N^l]^T 	\bm x_i) \notag \\
	&\quad + [b_1^l, b_2^l, \dots, b_N^l] \sin([w_1^l, w_2^l, \dots, w_N^l]^T \bm x_i) \notag \\
	&= A_0 + \bm{A}^l \cos(\mathbf{W}^l \bm x_i) + \bm{B}^l \sin(\mathbf{W}^l \bm x_i), \label{eq4}
\end{align}
where $[...]^T$ and $[...]$ denote the concatenation along the first and
second dimensions, respectively. To enrich the fitting possibilities of the neural network, the neural network embedded with Fourier series is extended and uses different angular frequency weight matrices $\bm{W}_{a}^l=[w_{1a}^l,w_{2a}^l,w_{3a}^l,\dots,w_{Na}^l]^T$, and $\bm{W}_{b}^l=[w_{1b}^l,w_{2b}^l,w_{3b}^l,\dots,w_{Nb}^l]^T$ , namely
\begin{align}
	f^{l}_\theta(\bm x_i) 
	&\approx A_0 + \bm{A}^l \cos(\mathbf{W}_{a}^l \bm x_i) + \bm{B}^l \sin(\mathbf{W}_{b}^l \bm x_i) \nonumber \\
	&= A_0 + \bm{\Theta}^l \left[ \cos(\mathbf{W}_a^l \bm x_i), \sin(\bm{W}_b^l \bm x_i) \right]^T \nonumber \\
	&= \mathcal{F}^l \circ (\bm x_i),
	\label{eq5}
\end{align}
where $\bm{\Theta}^l=[\bm{A}^l,\bm{B}^l]$ and $\mathcal{F}^{l}$ denotes the functional mapping of the $l$-th Fourier-basis layer. Inspired by Fourier analysis networks (FANs) \cite{dong2024fan}, the stacked Fourier layers and the Fourier-basis mapping can be expressed as 
\begin{equation}
	\begin{aligned}
		f_\theta(\bm{x}) 
		&= \mathcal{F}^{L}(\mathcal{F}^{L-1} \circ \mathcal{F}^{L-2} \circ \cdots \circ \mathcal{F}^{1} \circ \bm{x}) \\
		&= \bm{A}^L + \bm{\Theta}^{L} \left[ 
		\cos\left(\mathbf{W}_{\text{a}}^{L} (\mathcal{F}^{1:L-1} \circ \bm{x})\right) \parallel 
		\sin\left(\mathbf{W}_{\text{b}}^{L} (\mathcal{F}^{1:L-1} \circ \bm{x})\right) 
		\right],
	\end{aligned}
	\label{eq6}
\end{equation}
where $\bm{x}=[x_1,x_2, ..., x_i, ..., x_{d_x}]^T$, and $d_x$ is the dimension of $\bm{x}$. This results in very poor learning of the coefficient matrix $\bm{\Theta} ^ { L }$, and it would undergo more refined learning as the depth of the neural network increases. Therefore, in each layer, we follow the strategy of FAN \cite{dong2024fan} and parallelize the coefficient matrix $\bm{\Theta} ^ { l }$ and angular frequency matrix $\bm{W}_a^l\in \mathbb{R} ^ { d _  { p }\times d_x  }$  and  $ \bm{W}_b^l\in \mathbb{R} ^ { d _  { p }\times d_x  }$ in the same layer of neural network for learning, namely
\begin{equation}
	\phi ^{l}( \bm{x} ) \triangleq [ \cos ( \mathbf{W} _ { a }^l\bm{ x} ) , \sin ( \mathbf{W} _ { b } ^l\bm{x} ) , \sigma (\bm{A}_{c}^l+\bm{\Theta}_{c}^l\bm{x} ) ]^T,
	\label{eq7}
\end{equation}
where $\bm{A}_{c}^l \in \mathbb{R} ^ { d _  {\theta} }$ and $\bm{\Theta}_{c}^l\in \mathbb{R} ^ { d _  {\theta}\times d_x}$ are learnable parameters. Here, $d_\theta$ denotes the first dimension of $\bm{\Theta}_c$. We refer trigonometric nodes that perform cosine and sine operations as Fourier nodes. The fitting capability of a neural network is closely related to the diversity of basis functions at each node, therefore, two learnable bias matrices are added to the trigonometric operations of the Fourier nodes, namely, $\bm{B} _ {a} \in \mathbb{R} ^ { d _  { p }  },\bm{B} _ {b} \in \mathbb{R} ^ { d _  { p }  }$, where $d_p$ is half the number of Fourier neural nodes. Here, we divide the cosine and sine calculations equally. According to the trigonometric addition formulas, it can be concluded that the trigonometric function with added bias provides fitting possibilities for Fourier nodes such as $\text{cos} ^ 2$, $\text{sin} ^ 2$, and $\text{cos} \cdot \text{sin} $. Additional learnable coefficients $C_a$  and  $C_b \in \mathbb{R}$ are added at the Fourier nodes to increase the construction complexity of the neural network layers. Therefore, our final Fourier-basis layer is
\begin{equation}
	{{\bm y}_f^{l}}(\bm{x}) = \left[ C_a^l \cos(\mathbf{W}_a^l \bm{x} + \bm{B}_a^l) , C_b^l \sin(\mathbf{W}_b^l \bm{x} + \bm{B}_b^l) , \sigma(\bm{A}_c^l + \bm{\Theta}_c^l \bm{x}) \right]^T, \quad \text{if } l < L.
	\label{eq8}
\end{equation}

\subsection{\label{sec:level-PIWF}Physics-informed wavelet--Fourier model}

Fourier series are effective for representing the global modes that organize smooth advection, pressure variation, and wake-scale coherence, whereas wavelet series are naturally suited to localized high-frequency details such as sharp gradients, discontinuities, and compact vortical structures. Motivated by these complementary physical roles, the proposed physics-informed wavelet--Fourier model combines a Fourier-basis layer for global flow content with a wavelet layer for multiscale local flow content. These two layers are used in parallel with a conventional fully connected residual layer to form a three-branch representation.

As shown in Fig.~\ref{architecture}, to adaptively fuse the heterogeneous features from the three parallel branches, we employ a channel attention mechanism \cite{qin2021fcanet}. Treating the outputs of the wavelet, Fourier bases, and MLP layers as three distinct channels, we compute attention weights using a lightweight two-layer MLP and employ a channel attention mechanism to fuse the information from these three parallel branches. Given the concatenated features ${\bm Y}^{l} = [\bm y_w^{l},\bm y_f^{l},\bm y_m^{l}]\in\mathbb R^{3\times d}$, the attention score matrix $\bm S$ is obtained as 
\begin{equation}
	\bm S = \delta(\mathbf {W}_{\rm att, 2}\sigma(\bm {W}_{\rm att, 1}\bm Y)),
\end{equation}
where $\sigma$ denotes the GELU activation function \cite{lee2023mathematical}, $\mathbf W_{\rm att, 1}$ and $\mathbf W_{\rm att, 2}$ are learnable weight matrices. $\delta$ is a composite activation function, which will be presented in the following text.

Standard softmax normalization tends to produce overly sharp distributions and can suffer from numerical instability for large channel dimensions. To mitigate these issues, we adopt a scalable softmax followed by a sigmoid function defined by
\begin{equation}
	\delta(\cdot) = \sigma_{\rm{s}}(\sigma_{\rm{ssm}}(\cdot)),
\end{equation}
where $\sigma_{\rm{s}}(\bm{x}) = \frac{1}{1+e^{-\bm{x}}}$ is the sigmoid function. The scalable softmax function is defined as
\begin{equation}
	\sigma_{\rm{ssm}}:z _ { i } \mapsto \frac { n ^ { s z _ { i } } } { \sum _ { j = 1 } ^ { n } n ^ { s z _ { j } } } = \frac { e ^ { ( s \log n ) z _ { i } } } { \sum _ { j = 1 } ^ { n } e ^ { ( s \log n ) z _ { j } } },
\end{equation}
where $s$ is a scaling hyperparameter and $n=3$ denotes the number of branches. This combination yields a smoother and more balanced attention distribution that prevents any single branch from dominating while still emphasizing the most informative features.

Inspired by residual learning, the attention mechanism is applied in a residual form, namely
\begin{equation}
	\bm y^{l} = \bm {S } +\bm Y^{l} = \delta (\mathbf W _ {\rm att, 2 } \sigma ( \mathbf W _ {\rm att, 1} \bm Y^{l}))+\bm Y^{l}.
\end{equation}
With the global modeling capability now provided by the Fourier-basis layer, the wavelet layer can focus exclusively on detail spaces. Accordingly, the truncation strategy for the wavelet decomposition levels is adjusted so that $j$ starts from larger values, allowing the wavelet branch to concentrate on high-frequency localized features.

Finally, physical constraints are incorporated by embedding the governing partial differential equations, boundary conditions, and initial conditions directly into the loss function. The complete implementation strategy of the PIWF model is summarized in Algorithm~\ref{alg:PIWF_training}.
\begin{widetext}
\begin{center}
\begin{minipage}{0.94\textwidth}
\refstepcounter{algorithm}
\noindent\textbf{ALGORITHM~\thealgorithm.} Optimization of the PIWF model
\label{alg:PIWF_training}
\par\vspace{2mm}
\hrule
\vspace{1pt}
\hrule
\vspace{2mm}
{\small
\begin{algorithmic}[1]
\REQUIRE Initial data $\mathcal{D}_{\rm ic}$, boundary data $\mathcal{D}_{\rm bc}$, PDE data $\mathcal{D}_{\rm pde}$, 
loss weights ($\lambda_1,\lambda_2,\lambda_3$), maximum iterations $N_{\rm iter}$.
\ENSURE Trained PIWF parameters $\theta^{*}$.

\STATE Initialize all trainable parameters $\theta$.
\FOR{$n=1,\ldots,N_{\rm iter}$}

    \STATE Set the network input as $\bm y^{0}=(\bm x,t)$.

    \FOR{$l=1,\ldots,L-1$}
        \STATE Evaluate the three parallel branches at the $l$-th layer:
        \[
        \begin{aligned}
        &\text{Wavelet branch:}\quad
        \bm y_w^{l}
        =
        \begin{bmatrix}
        \mathbf W_{1}^{l}\Phi_w^{l}(\bm y^{l-1})\\
        \sigma(\mathbf B^{l}+\mathbf W_{2}^{l}\bm y^{l-1})
        \end{bmatrix},\\[-1mm]
        &\hspace{2.6cm}{\rm where}\quad
        \Phi_w^{l}(\bm y^{l-1})
        =
        [\phi_{J,0}(\bm y^{l-1}),\psi_0(\bm y^{l-1}),\ldots,
        \psi_{N-1}(\bm y^{l-1})]^{T},\\[1mm]
        &\text{Fourier-basis branch:}\quad
        \bm y_f^{l}
        =
        [C_a^l\cos(\mathbf W_a^l\bm y^{l-1}+\bm B_a^l),\\[-1mm]
        &\hspace{4.15cm}
        C_b^l\sin(\mathbf W_b^l\bm y^{l-1}+\bm B_b^l),
        \sigma(\bm A_c^l+\bm\Theta_c^l\bm y^{l-1})]^{T},\\[1mm]
        &\text{MLP branch:}\quad
        \bm y_m^{l}
        =
        \sigma(\mathbf W_m^{l}\bm y^{l-1}+\bm b_m^{l}).
        \end{aligned}
        \]

        \STATE Fuse the three branches through residual channel attention:
        \[
        \bm Y^{l}
        =
        [\bm y_w^{l},\bm y_f^{l},\bm y_m^{l}]
        \in\mathbb R^{3\times d},
        \;\; {\rm and}\;\;
        \bm y^{l}
        =
        \bm Y^{l}
        +
        \delta\!\left(
        \mathbf W_{{\rm att},2}^{l}
        \sigma(\mathbf W_{{\rm att},1}^{l}\bm Y^{l})
        \right).
        \]
    \ENDFOR

    \STATE Obtain the PIWF prediction from the final-layer representation:
    $
    \bm u_\theta(\bm x,t)
    =
    \mathbf W_{L}\bm y^{L}
    +
    \bm b_{L}.
    $

    \STATE Compute the PDE residual by automatic differentiation:
    $
    \mathcal L_{\rm pde}=
    \partial_t\bm u_\theta
    -
    R(\bm u_\theta;\bm x,t).
    $

    \STATE Evaluate $\mathcal L_{\rm ic}$, $\mathcal L_{\rm bc}$, and $\mathcal L_{\rm pde}$ on 
    $\mathcal{D}_{\rm ic}$, $\mathcal{D}_{\rm bc}$, and $\mathcal{D}_{\rm pde}$.

    \STATE Update $\theta$ by minimizing
    $
    \mathcal L(\theta)
    =
    \mathcal L_{\rm ic}(\theta)
    +
    \mathcal L_{\rm bc}(\theta)
    +
    \mathcal L_{\rm pde}(\theta).
    $

\ENDFOR
\RETURN $\theta^{*}$.
\end{algorithmic}
}
\vspace{2mm}
\hrule
\vspace{1pt}
\hrule
\vspace{2mm}
\end{minipage}
\end{center}
\end{widetext}

The PIWF framework is implemented as a modular physics-informed learning package in Python. PyTorch \cite{paszke2019pytorch} is employed to construct the neural-network architecture and perform automatic differentiation, whereas DeepXDE \cite{lu2021deepxde} is used to manage the physics-informed training procedure, including the construction of residual losses and the enforcement of initial and boundary conditions. The main components of PIWF, including the wavelet-series feature layer, Fourier-basis feature layer, residual neural network, and channel-attention module, are encapsulated as standard \texttt{nn.Module} classes. This modular implementation facilitates code reuse, ablation studies, and extensions to different partial differential equations. For a given set of spatio-temporal coordinates, the wavelet branch extracts localized multiscale representations, and the Fourier branch provides global spectral features. These features are combined with the residual-network output and subsequently processed by a channel-attention module to obtain the final prediction of the physical variables. The residuals of the governing equations are evaluated by automatic differentiation, with the required derivatives computed directly from the differentiable computational graph. The complete loss function, consisting of equation residuals together with initial and boundary condition penalties, is then minimized in an end-to-end training process. The implementation is fully compatible with GPU acceleration through the CUDA backend. All major tensor operations, basis evaluations, feature transformations, and loss evaluations are performed within the PyTorch computational graph, thereby reducing unnecessary host-device data transfer and ensuring stable gradient propagation. This software design provides an efficient and reproducible implementation of the proposed PIWF method for solving multiscale nonlinear partial differential equations.

\section{\label{sec:level1/3} Results}
This section evaluates PIWF on five fluid-dynamics problems chosen to isolate different physical mechanisms: viscous shock formation in Burgers' equation, wet--dry-front propagation in the shallow-water equations, steady viscous wake recovery in Kovasznay flow, decaying vortical motion in Taylor--Green vortex flow, and coherent vortex shedding in the wake of a circular cylinder. The proposed model is compared with a high-resolution finite-difference (FD) solver where appropriate, a standard physics-informed neural network (PINN), and a physics-informed Kolmogorov--Arnold network (PIKAN; see Appendix~\ref{sec:PIKAN}). All neural-network models are trained without high-fidelity interior reference data, using boundary and initial conditions when present together with the governing-equation residual. High-fidelity reference fields are used only for quantitative assessment. This design allows the comparison to focus on whether the learned solution preserves the relevant flow physics under sparse constraints. Unless otherwise stated, the Fourier node ratio is $p_{\rm{ratio}}=0.15$ and the wavelet node ratio is $q_{\rm{ratio}}=0.3$. The channel-attention module uses a two-layer MLP with 256 neurons per layer and GELU activation, and all models are trained with the Adam optimizer at a fixed learning rate of 0.001 on an RTX 4090 GPU.

\subsection{Burgers' equation}

The Burgers' equation is given by
\begin{equation}
	\frac{\partial u}{\partial t}+u \frac{\partial u}{\partial x}={\nu} \frac{\partial^{2} u}{\partial x^{2}},   \quad        \mathrm{where} \;\; x \in[-1,1], \;{\rm and}\;\; t \in[0,1].
\end{equation}
Here, $u(x,t)$ represents the fluid velocity, $\nu = {1}/{\rm Re} $ represents viscosity coefficient and ${\rm Re}$ is the Reynolds number. The equation is supplemented with the Dirichlet boundary conditions $u(-1,t)=u(1,t)=0$ and the initial condition $u(x,0)=-\sin(\pi x)$.

To examine different levels of nonlinear steepening, we consider two representative Reynolds numbers, ${\rm Re}=100\pi$ and $1000$. In the lower-Reynolds-number case, viscous diffusion still regularizes the convectively generated gradient. In the higher-Reynolds-number case, the velocity field develops a much thinner shock-like layer whose position and amplitude are sensitive indicators of the balance between nonlinear convection and viscous dissipation. This case is particularly demanding for neural solvers because an accurate prediction must localize the steep transition without introducing spurious oscillations or excessive numerical smoothing.

The ground-truth solution for quantitative comparison is obtained via the Cole-Hopf transformation \cite{cite-key}, which converts the nonlinear Burgers' equation into the linear heat equation. Specifically, the transformation $u = -2\nu 
\partial (\ln\phi)/\partial x$ converts the Burgers' equation into the heat equation ${\partial \phi}/{\partial t} = \nu {\partial^2 \phi}/{\partial x^2}$. The analytical solution of the heat equation can be expressed as the ratio of two infinite integrals. In the numerical implementation, we employ the Sobol low-discrepancy sequence to generate uniformly distributed sampling points within the integration interval and use the Quasi-Monte Carlo method \cite{hickernell2025quasimontecarlomethodswhat} to compute the two integrals efficiently, thereby obtaining the exact solution at any spatio-temporal point $(x,t)$. This approach provides higher accuracy than conventional finite difference methods and is therefore suitable as a benchmark solution for validating numerical methods.

A high-resolution finite difference solver is employed as a classical numerical baseline.
The solver adopts a second-order central difference scheme for spatial discretization and the forward Euler method for time marching. To ensure stability and satisfy the CFL condition, a significantly denser grid than that used for the neural network collocation points is employed. Specifically, for the low- and high-Reynolds-number cases, the spatial domain $x \in [-1,1]$ is discretized into 1000 and 40000 grid points, respectively, and the temporal interval $t \in [0,1]$ is divided into 2000 and 3000 time steps, respectively. The same initial condition $u(x,0)=-\sin(\pi x)$ and Dirichlet boundary conditions $u(-1, t) = u(1, t) = 0$ are imposed. At each time step, the first and second spatial derivatives $u_x$ and $u_{xx}$ are computed using the second-order central difference scheme at interior points, and the solution is advanced explicitly via the forward Euler method, namely $u^{n+1} = u^n + \Delta t (\nu u_{xx} - u^n u_x)$.

For the low-Reynolds-number case (${\nu}$=0.01/$\pi$), the PINN, PIKAN, and PIWF models are trained using 50 collocation points randomly sampled via Latin hypercube sampling \cite{Latin} on the boundaries for the Dirichlet conditions and 50 points at $t=0$ for the initial condition. The weight of the PDE residual term in Eq. \ref{16} is set to $\lambda_3=10^{-4}$. For the high-Reynolds-number case (${\nu}$=0.001), larger PDE residuals arise in the sharp shock region, potentially overwhelming the contributions from the initial and boundary conditions. To maintain training stability, the residual weight is reduced by one order of magnitude to $\lambda_3=10^{-5}$, while the numbers of boundary and initial condition points are both increased to 100.

Within the computational domain, 1000 high-fidelity reference data points are randomly sampled for performance evaluation and are strictly excluded from the training process. To enforce the PDE residual $L_{\rm{pde}}(\theta)$, 10000 collocation points are uniformly distributed across the spatio-temporal domain. 
All neural networks (PINN, PIKAN, and PIWF) are designed to reconstruct the entire flow field by enforcing physical constraints through the residual loss, using only the given initial and boundary conditions. The baseline PINN employs a fully connected MLP with 6 hidden layers and 50 neurons per hidden layer. The PIKAN architecture replaces the MLP backbone with a KAN of the same depth and hidden dimension, i.e., 6 hidden layers with a hidden dimension of 50. In contrast to MLPs, where nonlinear activations are applied at neurons, KAN parameterizes learnable univariate functions on the edges. For PIWF, each of the three parallel branches adopts the same configuration as the baseline PINN, with 6 hidden layers and 50 neurons per layer, while no direct connections are introduced among different branches. The outputs of these branches are finally fused by a lightweight channel-attention module, implemented using a two-layer MLP with 256 neurons per hidden layer, which adaptively assigns channel-wise weights to balance the contributions of different feature representations.
 The hyperbolic tangent activation function is employed in the hidden layers. The networks map the spatio-temporal coordinates $(x,t)$ to the velocity field $u$, resulting in input and output layers with two and one neurons, respectively. All models are trained for 200000 epochs using the Adam optimizer to ensure adequate convergence of the loss function.

\begin{figure}[H]
	\centering
	\includegraphics[width=0.9\textwidth]{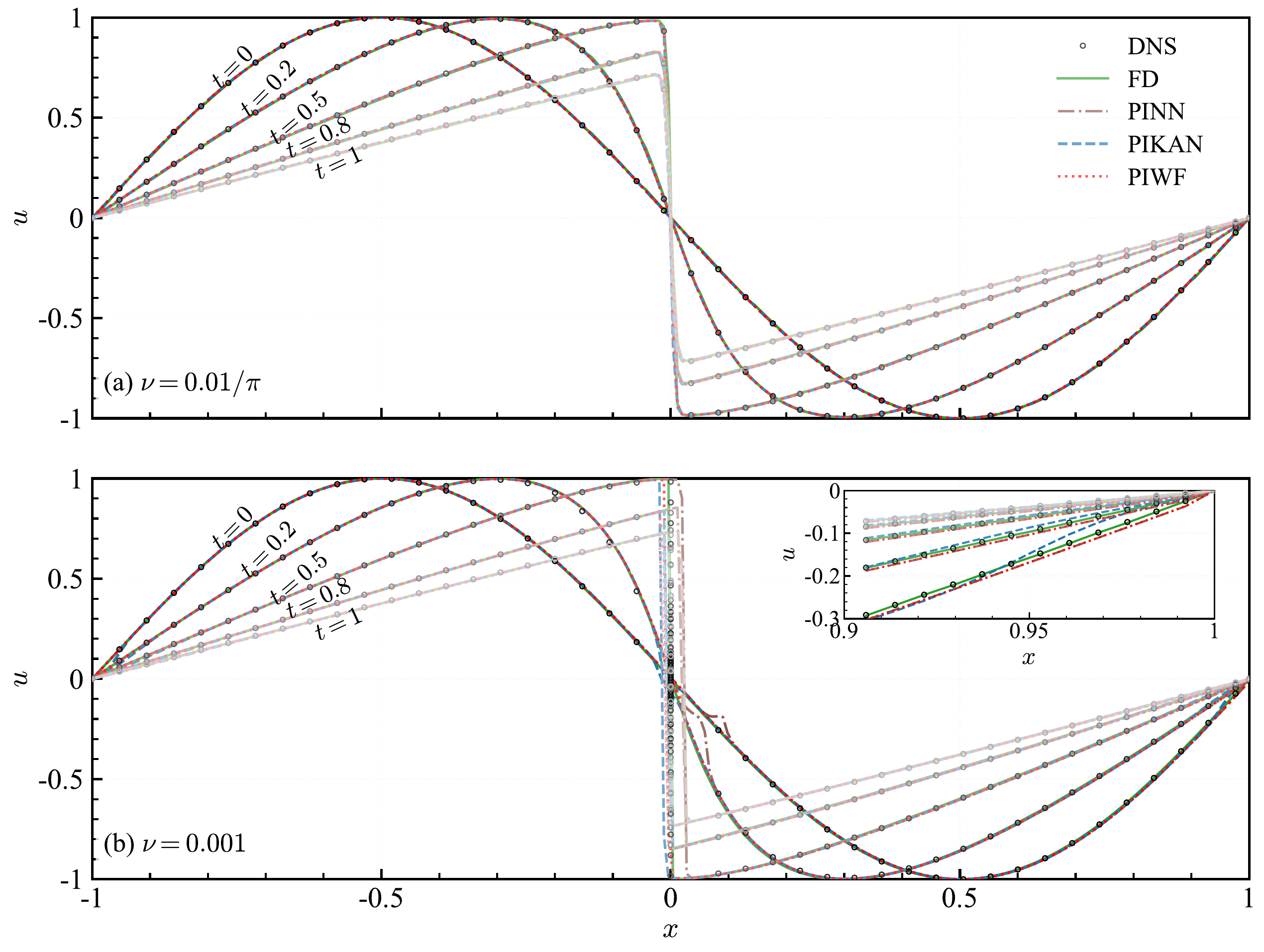}
	\caption{Comparison of velocity profiles predicted by the FD, PINN, PIKAN, and PIWF models for the Burgers' equation at two different viscosities. (a) low-Reynolds-number case with $\nu=0.01/\pi$; (b) high-Reynolds-number case with $\nu = 0.001$.}
	\label{burgerspred}
\end{figure}

Fig. \ref{burgerspred} compares the predicted velocity profiles from the FD, PINN, PIKAN, and PIWF models at representative time instants. The comparison highlights the ability of each method to preserve the physically localized shock layer. In the lower-Reynolds-number case, all models reproduce the overall velocity evolution, but differences appear near the steepest part of the profile. In the higher-Reynolds-number case, these differences become more pronounced: PIWF follows the reference shock front most closely, whereas the standard PINN produces larger deviations both at the discontinuity and in its immediate neighborhood.

\begin{figure}[htbp] 
	\centering
	\includegraphics[width=0.90\textwidth]{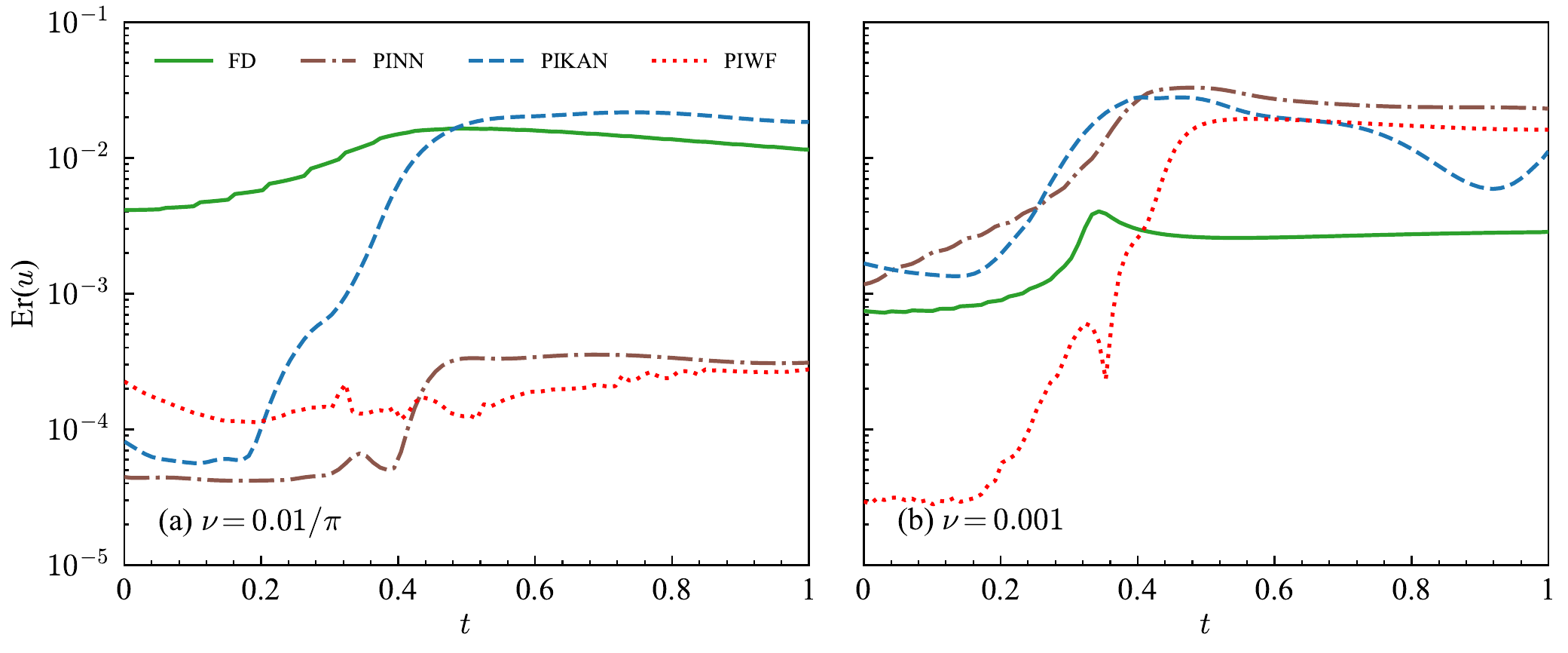}
	\caption{Relative error evolution along time $t$ for the Burgers' equation predicted by the FD, PINN, PIKAN, and PIWF models. (a) low-Reynolds-number case with $\nu = 0.01 / \pi$; (b) high-Reynolds-number case with $\nu = 0.001$. }
	\label{burgerser1}
\end{figure}

To quantitatively examine the predictive performance of each model, the temporal evolution of the relative error is illustrated in Fig. \ref{burgerser1}. The root-mean-square relative error is defined by
\begin{equation}
	Er_{\text{rms}} = \frac{\sqrt{\frac{1}{N} \sum_{i=1}^{N} (u_{\text{true}, i} - u_{\text{pred}, i})^2}}{\sqrt{\frac{1}{N} \sum_{i=1}^{N} (u_{\text{true}, i})^2}} .
\end{equation}
In both low- and high-Reynolds-number cases, the finite difference method exhibits non-negligible errors despite using a mesh that is more than two orders of magnitude denser than the collocation points employed by the neural-network models.

In the low-Reynolds-number case, both PINN and PIKAN models exhibit a pronounced localized increase in error near the discontinuity. The PIKAN model demonstrates the poorest performance, with its error in the high-gradient region even exceeding that of the finite difference method. In contrast, the proposed PIWF effectively suppresses error amplification in the steep-gradient zone and achieves the lowest error magnitude near the discontinuity. However, its accuracy in the smoother, low-gradient regions is slightly lower than that of PINN and PIKAN.

This behavior reflects the role of the wavelet basis in concentrating representational capacity around the physical region where the velocity gradient is largest. At equivalent computational cost, PIWF preferentially resolves the shock layer, where small phase or amplitude errors have a disproportionate influence on the solution. Because the optimizer minimizes a global loss, improved shock localization can be accompanied by a slight redistribution of error in smoother low-gradient regions.

A related tendency is observed for PIKAN in the high-Reynolds-number regime. As the velocity gradient steepens, the model devotes more capacity to the discontinuity, but the resulting representation is less successful at maintaining the surrounding smooth field. By contrast, PIWF balances global Fourier content and localized wavelet corrections, thereby preserving the shock while keeping the low-gradient regions relatively stable.

\begin{figure}[H]
	\centering
	\includegraphics[width=0.9\textwidth]{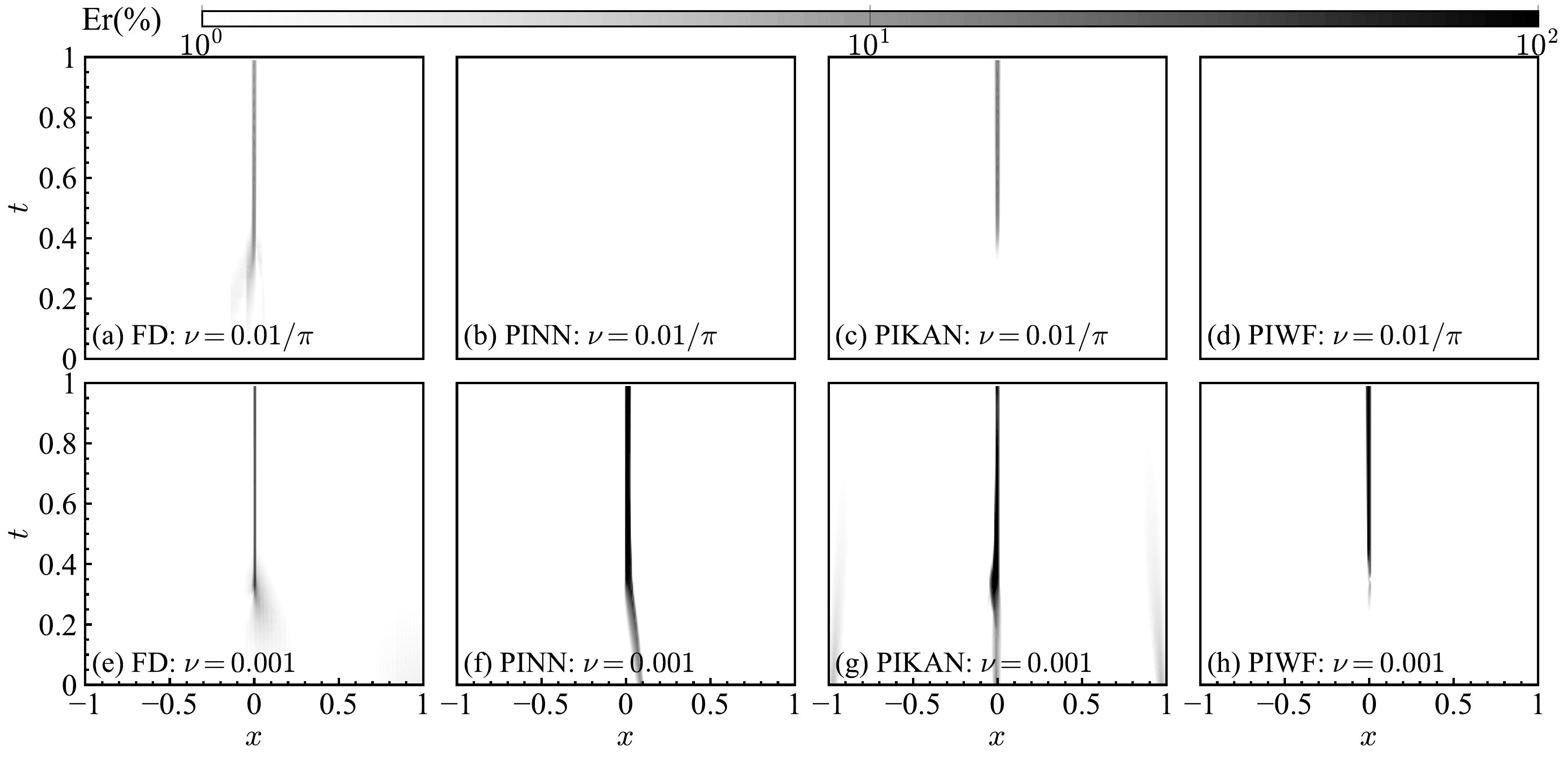}
	\caption{Relative error contours for the Burgers' equation predicted by the FD, PINN, PIKAN, and PIWF models at two different viscosities. The top row corresponds to the low-Reynolds-number case with $\nu=0.01/\pi$, and the bottom row corresponds to the high-Reynolds-number case with $\nu = 0.001$.}
	\label{bgcontour1}
\end{figure}

Fig. \ref{bgcontour1} illustrates the relative error contours for the Burgers' equation predicted by the FD, PINN, PIKAN, and PIWF models. 
In the low-Reynolds-number case, the standard PINN achieves overall satisfactory performance. Lacking embedded basis functions, it behaves similarly to a global approximator: the network layers learn the solution through spatially averaged representation governed by the universal approximation theorem. Although PIKAN and PIWF share the idea of embedding basis-function representations into hidden-layer nodal operations, they differ in the choice of bases: PIKAN relies on B-splines, whereas PIWF combines wavelet and Fourier series to enhance multiscale representation. While this design allows the neural networks to allocate more representational capacity to challenging shock regions, the global nature of the optimization often forces a rebalancing of previously converged smooth regions, introducing additional errors in low-gradient areas.

Although both PIWF and PIKAN are designed to enhance the representation of localized features, PIWF achieves significantly better shock resolution than the standard PINN, whereas PIKAN still exhibits substantially larger errors near the discontinuity. This performance gap arises primarily from the inherent limitations of B-splines in representing discontinuous solutions. B-spline functions are optimized for smooth and continuous manifolds. When faced with sharp gradients or abrupt transitions, they tend to enforce global regularity at the expense of local fidelity, resulting in artificial smoothing of the shock and loss of high-frequency detail.

Even with sufficient computational resources, the PIKAN model still faces difficulty in concentrating its learning capacity on localized complex regions when regions of substantially different complexity have to be resolved simultaneously. The large PDE residuals generated by steep velocity gradients tend to dominate the composite loss function, causing the optimizer to overemphasize  the shock region. This often leads to marginalization of smoother zones and, in some cases, compromises the satisfaction of initial and boundary conditions. 

This issue is particularly acute for architectures that embed basis functions directly into the network nodes, as their primary goal is to enhance localized feature extraction. In such models, balancing contributions across regions is difficult since all nodal weights are coupled under the universal approximation theorem. In contrast, the proposed PIWF effectively modulates the relative importance of global and local features through its channel attention mechanism, thereby avoiding the pronounced degradation in boundary and initial condition accuracy observed in other basis-embedded frameworks.

\subsection{Shallow water equations}
The shallow water equations (SWEs) are a system of nonlinear, time-dependent, hyperbolic partial differential equations that describe the water depth and horizontal momentum of free-surface flows. They can be expressed as
\begin{align}
	\frac{\partial h}{\partial t} + \frac{\partial (h u)}{\partial x} &= 0, \label{eq:continuity} \\
	\frac{\partial (h u)}{\partial t} + \frac{\partial \left( h u^{2} + g h^{2}/2 \right)}{\partial x} &= g h (s_{ox} - s_{fx}). \label{eq:momentum}
\end{align}
Here, $h(x,t)$ is the water depth, $u(x,t)$ is the depth-averaged velocity, and $g$ is the gravitational acceleration. The term $s_{ox}=-\partial z/\partial x$ denotes the bed slopes, with $z(x)$ being the time-invariant topographic elevation. The bed friction slope $s_{fx}$ is modeled using the Manning--Strickler formula \cite{yen1992dimensionally}, namely
\begin{equation}
	s_{fx} = n^{2} u \sqrt{u^{2} + v^{2}} \, h^{-\frac{4}{3}}, 
\end{equation}
where $n$ is the Manning coefficient. In this study, the solution is obtained by solving the SWEs in a fully physics-informed manner without high-fidelity reference data.

This test case considers wave propagation over a horizontal bed driven by a time-dependent boundary condition, and can be solved analytically \cite{hunter2005adaptive}
\begin{equation}
	h(x,t)=[\frac{7}{3}(n^{2}u^{2}(x-ut))]^\frac{3}{7},
	\label{ha}
\end{equation}
where the constant velocity is $u=0.29\;\mathrm{m}/\mathrm{s}$ and the Manning coefficient is $n=0.03\;\text{s} \text{m}^{-{1}/{3}}$. The inflow boundary condition at $x=0$ is given by Eq. \ref{ha}, while the domain is initialized by $h(x, 0) = 0$. 

The shallow water equations are particularly challenging for standard neural solvers because the wet--dry front is a moving free-boundary-like structure where the water depth changes abruptly and the discharge must remain physically consistent with the nearly dry state. In this study, this problem is used to evaluate whether PIWF can preserve both the front location and the coupled depth--momentum response without relying on high-fidelity interior reference data.

The computational domain for this numerical experiment is defined as $x \in [0, 1200]\;$m and $t \in [0, 3600]\;$s. The reference solution is given by the analytical expression in Eq. \ref{ha}. For training, the initial condition at $t=0$ is enforced using 121 uniformly spaced points with $\Delta x=10\;$m. The time-dependent inflow boundary condition at $x=0$ is imposed with a temporal resolution of $\Delta t=180\;$s. Model performance is evaluated on a test set consisting of a $121 \times 61$ grid. The PDE residual is enforced at 2541 collocation points arranged on a 
$121 \times 21$ grid. 

All neural architectures (PINN, PIKAN, and PIWF) consist of 6 hidden layers with 50 neurons per layer. The input vector ${\mathbf \mathit X}$ comprises the spatio-temporal coordinates $(x, t)$, the topographic slope, and the Manning friction coefficient, resulting in 4 input variables. The output vector ${\mathbf \mathit U}$ contains the water depth $h$ and discharge $hu$, yielding 2 output variables. All models are trained for 50000 epochs using the Adam optimizer.

For comparison, a high-resolution finite-difference solver is employed. The convective term is discretized using the first-order upwind scheme. The spatial domain is discretized with $\Delta x=1\;$m (1201 grid points), and the temporal domain uses $\Delta t=1\;$s (3601 time steps) to strictly satisfy the Courant-Friedrichs-Lewy (CFL) condition.

\begin{figure}[htbp] 
	\centering
	\includegraphics[width=0.9\textwidth]{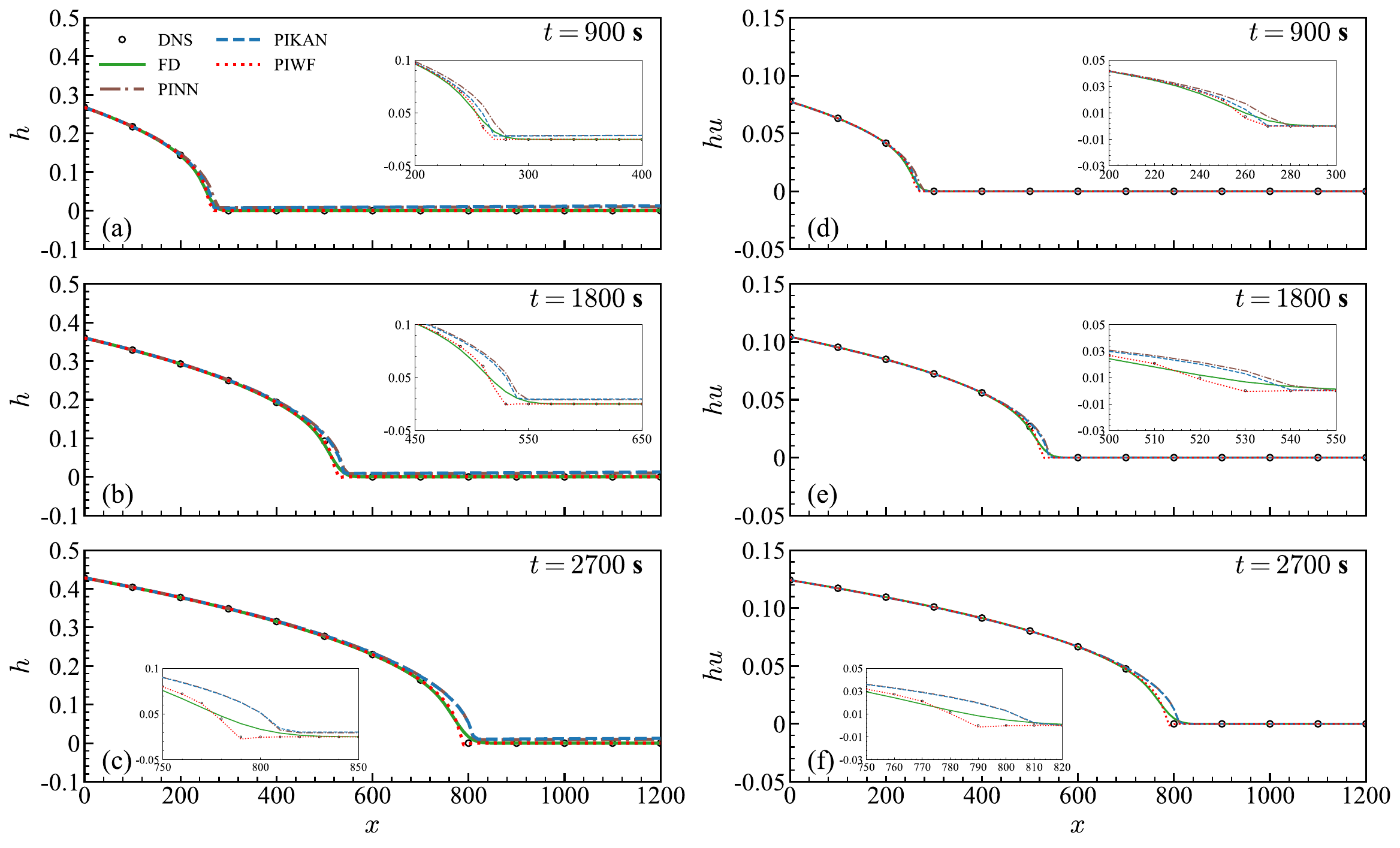}
	\caption{Comparisons of water depth $h$ and discharge $hu$ profiles predicted by the FD, PINN, PIKAN, and PIWF models for the shallow water equations at $t = 900$, $1800$, and $2700\;$s. Panels (a-c) and (d-f) show the spatial distributions of $h$ and $hu$, respectively. Magnified views near the wet-dry fronts are provided in the insets to highlight resolution quality.}
	\label{swehpred}
\end{figure}

Fig. \ref{swehpred} illustrates the predicted water depth $h$ and discharge $hu$ profiles along the $x$-direction at $t = 900$, $1800$, and $2700\;$s. The dominant physical feature is the advancing wet--dry front. PIWF provides the most robust reconstruction in the immediate vicinity of this front, maintaining consistency for both water depth and discharge even when the transition becomes extremely steep. The finite-difference solution is affected by the numerical dissipation of the first-order upwind scheme, which introduces artificial viscosity and smooths the ideally sharp front. PIKAN shows a modest improvement over the standard PINN near the front at $t = 900\;$s, indicating improved local representation, but its accuracy deteriorates more noticeably away from the front.
\begin{table}
	\centering
	\footnotesize
	\setlength{\tabcolsep}{4pt}
	\caption{Relative $L^2$ errors for the shallow water equations at different spatial locations. }
	\renewcommand{\arraystretch}{1.1}
	\resizebox{\textwidth}{!}{%
	\begin{tabular}{lcccccccc}
		\toprule
		Model / Position & $x=$0 & 200 & 400 & 600 & 800 & 1000 & 1200 & Average \\
		\midrule
		$h$ (FD)    & $5.066 \times 10^{-6}$ & $3.039 \times 10^{-3}$ & $6.538 \times 10^{-3}$ & $7.282 \times 10^{-3}$ & $9.660 \times 10^{-3}$ & $1.300 \times 10^{-2}$ & $1.265 \times 10^{-2}$ & $3.866 \times 10^{-3}$ \\
		$h$ (PINN)  & $4.981 \times 10^{-5}$ & $4.068 \times 10^{-2}$ & $4.517 \times 10^{-2}$ & $4.829 \times 10^{-2}$ & $5.347 \times 10^{-2}$ & $5.989 \times 10^{-2}$ & $6.344 \times 10^{-2}$ & $3.685 \times 10^{-2}$ \\
		$h$ (PIKAN) & $2.768 \times 10^{-4}$ & $3.412 \times 10^{-2}$ & $4.191 \times 10^{-2}$ & $4.781 \times 10^{-2}$ & $5.510 \times 10^{-2}$ & $6.334 \times 10^{-2}$ & $6.888 \times 10^{-2}$ & $3.935 \times 10^{-2}$ \\
		$h$ (PIWF)  & $8.499 \times 10^{-5}$ & $5.009 \times 10^{-4}$ & $1.494 \times 10^{-3}$ & $8.833 \times 10^{-4}$ & $1.603 \times 10^{-3}$ & $1.947 \times 10^{-3}$ & $2.357 \times 10^{-3}$ & $8.779 \times 10^{-4}$ \\
		\midrule
		$hu$ (FD)    & $8.304 \times 10^{-8}$ & $3.039 \times 10^{-3}$ & $6.538 \times 10^{-3}$ & $7.282 \times 10^{-3}$ & $9.660 \times 10^{-3}$ & $1.300 \times 10^{-2}$ & $1.265 \times 10^{-2}$ & $3.866 \times 10^{-3}$ \\
		$hu$ (PINN)  & $1.953 \times 10^{-6}$ & $5.378 \times 10^{-3}$ & $1.074 \times 10^{-2}$ & $1.512 \times 10^{-2}$ & $2.185 \times 10^{-2}$ & $3.021 \times 10^{-2}$ & $3.599 \times 10^{-2}$ & $9.899 \times 10^{-3}$ \\
		$hu$ (PIKAN) & $1.024 \times 10^{-4}$ & $1.184 \times 10^{-3}$ & $6.158 \times 10^{-3}$ & $1.147 \times 10^{-2}$ & $1.974 \times 10^{-2}$ & $2.968 \times 10^{-2}$ & $3.733 \times 10^{-2}$ & $8.405 \times 10^{-3}$ \\
		$hu$ (PIWF)  & $8.652 \times 10^{-5}$ & $4.557 \times 10^{-4}$ & $1.415 \times 10^{-3}$ & $6.686 \times 10^{-4}$ & $1.373 \times 10^{-3}$ & $1.721 \times 10^{-3}$ & $2.155 \times 10^{-3}$ & $8.742 \times 10^{-4}$ \\
		\bottomrule
	\end{tabular}%
	}
	\label{tab:table2}
\end{table}

Table \ref{tab:table2} summarizes the global relative $L^2$ errors for both $h$ and $hu$. Although PIKAN exhibits higher local accuracy near the wet-dry front, its overall error for water depth remains slightly higher than that of PINN. This indicates that PIKAN exhibits larger errors in the smoother and computationally simpler regions, particularly in the stabilized wet and dry zones. Among all tested frameworks, the PIWF model consistently delivers the lowest errors across both fields and all evaluation metrics.

\begin{figure}[htbp] 
	\centering
	\includegraphics[width=0.9\textwidth]{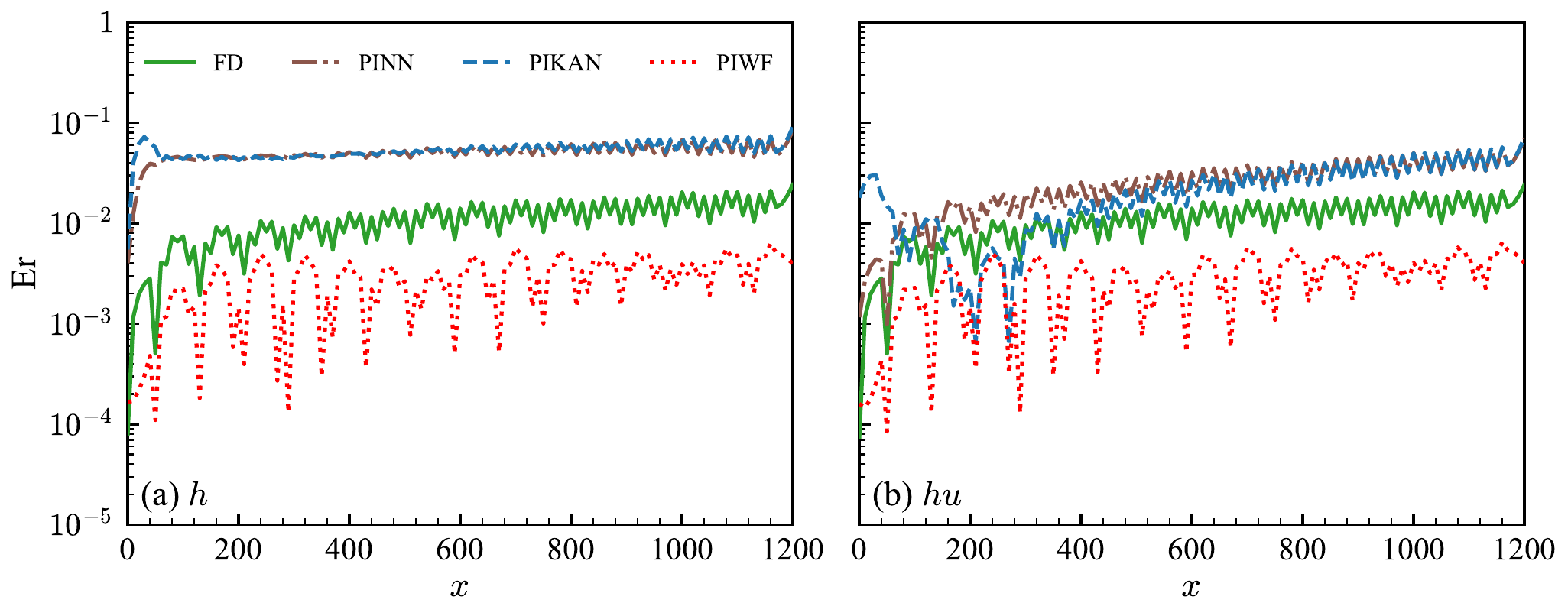}
	\caption{Relative $L^2$ error distributions along the $x$-direction for water depth $h$ and discharge $hu$ predicted by the FD, PINN, PIKAN, and PIWF models for the shallow water equations. (a) water depth $h$; (b) discharge $hu$.}
	\label{swepreder}
\end{figure}

Fig. \ref{swepreder} presents the relative $L^2$ error profiles along the $x$-direction for the water depth $h$ and discharge $hu$. The relative error exhibits noticeable oscillatory behavior for all models. This behavior can be primarily attributed to the presence of the wet-dry front, where the solution magnitude becomes very small, causing even minor prediction deviations to be significantly amplified in the relative-error measure. The finite-difference solution generally exhibits lower relative errors than PINN and PIKAN, particularly in the dry region. This indicates that the FD scheme maintains values closer to the near-zero reference state in these regions, although this behavior is partly associated with its dissipative smoothing of the wet-dry transition.

\begin{figure}[H]
	\centering
	\includegraphics[width=0.9\textwidth]{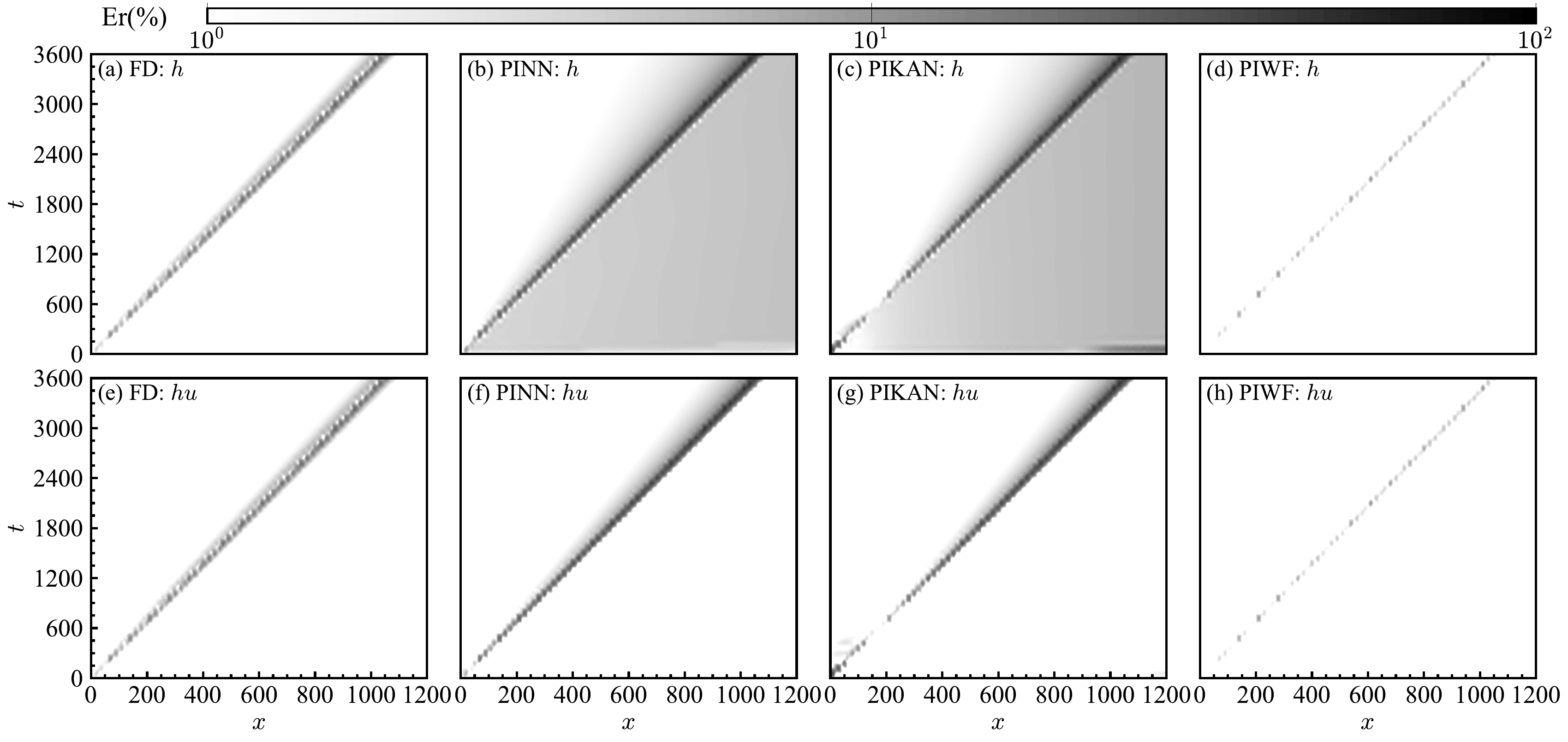}
	\caption{Relative error contours of water depth $h$ and discharge $hu$ predicted by the FD, PINN, PIKAN, and PIWF models for the shallow water equations.}
	\label{swecontour}
\end{figure}

Among all tested methods, PIWF achieves the lowest overall relative errors for both $h$ and $hu$. This improvement is also reflected in the relative error contours shown in Fig. \ref{swecontour}, where the error concentration around the wet--dry interface is weaker and more localized. Physically, this indicates that the model captures not only the front position but also the coupled momentum response across the transition. The compact support and orthogonality of the Haar wavelet basis allow the correction associated with the front to remain localized, while the Fourier branch preserves the smoother background propagation. This combination is particularly useful for discontinuity-dominated shallow-water dynamics.

\subsection{Kovasznay flow}\label{sec:kovasznay}

We next consider Kovasznay flow, an exact steady solution of the two-dimensional incompressible Navier--Stokes equations. This problem complements the preceding unsteady and discontinuity-dominated examples because it contains a smooth but slowly decaying viscous wake with coupled velocity and pressure fields. It therefore provides a controlled benchmark for examining whether a physics-informed model can recover a two-dimensional wake structure without using high-fidelity interior reference data. The governing equations are
\begin{equation}
	\begin{aligned}
		\frac{\partial u}{\partial x}+\frac{\partial v}{\partial y} &= 0, \\
		u\frac{\partial u}{\partial x}+v\frac{\partial u}{\partial y}
		&= -\frac{\partial p}{\partial x}
		+\frac{1}{\rm Re}\left(\frac{\partial^2 u}{\partial x^2}+\frac{\partial^2 u}{\partial y^2}\right),\\
		u\frac{\partial v}{\partial x}+v\frac{\partial v}{\partial y}
		&= -\frac{\partial p}{\partial y}
		+\frac{1}{\rm Re}\left(\frac{\partial^2 v}{\partial x^2}+\frac{\partial^2 v}{\partial y^2}\right).
	\end{aligned}
	\label{eq:kov-ns}
\end{equation}
The analytical solution is written as
\begin{equation}
	\begin{aligned}
		u(x,y) &= 1-\exp(\lambda x)\cos(2\pi y),\\
		v(x,y) &= \frac{\lambda}{2\pi}\exp(\lambda x)\sin(2\pi y),\\
		p(x,y) &= \frac{1}{2}\left[1-\exp(2\lambda x)\right],
	\end{aligned}
	\label{eq:kov-solution}
\end{equation}
\begin{equation}
	\lambda=\frac{\rm Re}{2}-\sqrt{\frac{{\rm Re}^2}{4}+4\pi^2}.
	\label{eq:kov-lambda}
\end{equation}
The Reynolds number is set to ${\rm Re}=1000$, and the computational domain is $(x,y)\in[-0.5,40]\times[-0.5,1.5]$. The extended streamwise length is used because, at this Reynolds number, the exponential decay rate is small and the wake structure persists over a long distance. Dirichlet boundary values are prescribed from Eq.~\ref*{eq:kov-solution} on all four sides of the domain. The training set contains 2601 interior collocation points and 101 points on each boundary segment, while an independent set of 400 analytical points is used only for testing. For this steady two-dimensional case, the input vector is $(x,y)$ and the model output is $(u,v,p)$. PINN, PIKAN, and PIWF use the same depth and width, with four hidden layers and 50 neurons per hidden layer, and all models are trained for 400000 epochs.

\begin{figure}[H]
	\centering
	\includegraphics[width=0.9\textwidth]{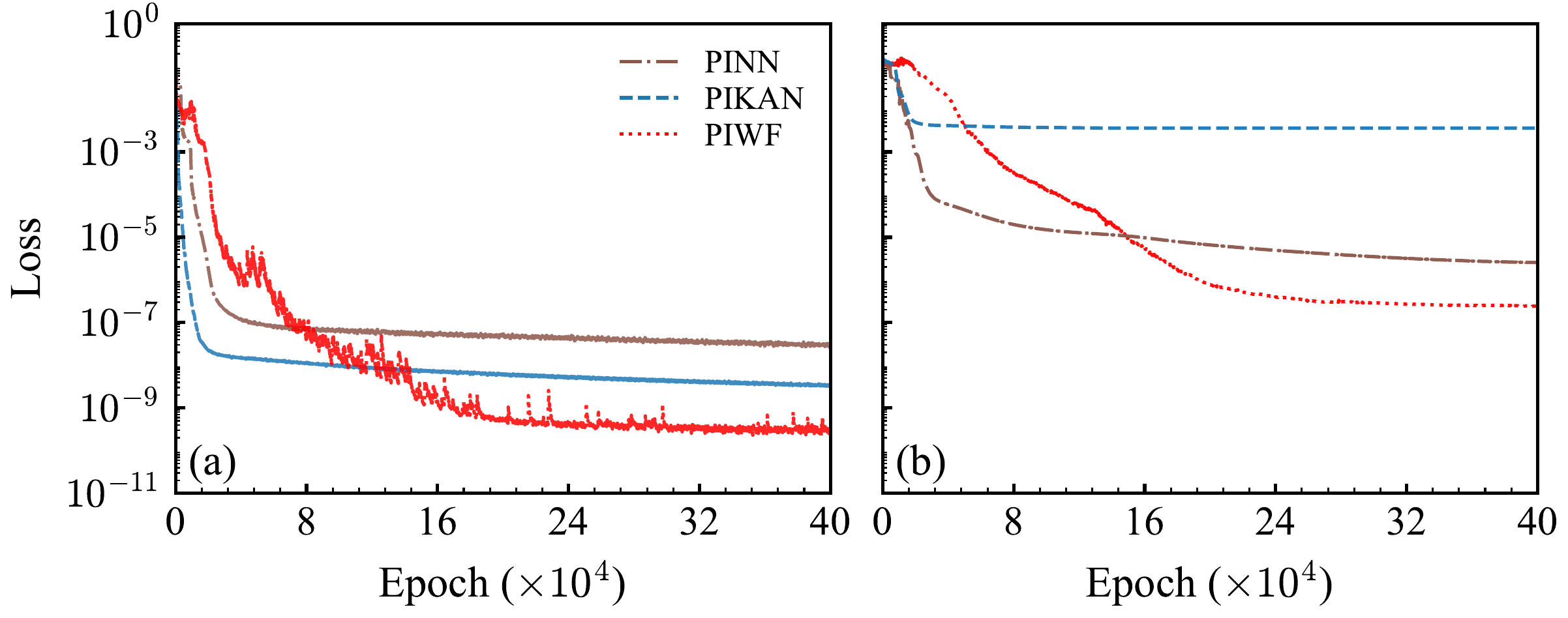}
			\caption{Learning curves of the PINN, PIKAN, and PIWF models for Kovasznay flow at ${\rm Re}=1000$: (a) training loss; (b) testing loss.}
	\label{kovloss}
\end{figure}

The learning curves are shown in Fig.~\ref*{kovloss}. The three models all reduce the physics-informed objective during training, but their testing behavior differs markedly. PIKAN reaches a nearly stagnant testing loss after the early stage, indicating that the spline-based representation has difficulty improving the long downstream wake once the leading-order field has been fitted. PINN continues to reduce the testing error, whereas PIWF obtains the lowest final testing loss. This behavior is consistent with the structure of the exact solution: the Fourier branch represents the smooth periodic dependence in the transverse direction, while the wavelet branch provides localized corrections where the wake amplitude and pressure gradient change most rapidly.

\begin{figure}[htbp]
	\centering
	\includegraphics[width=0.9\textwidth]{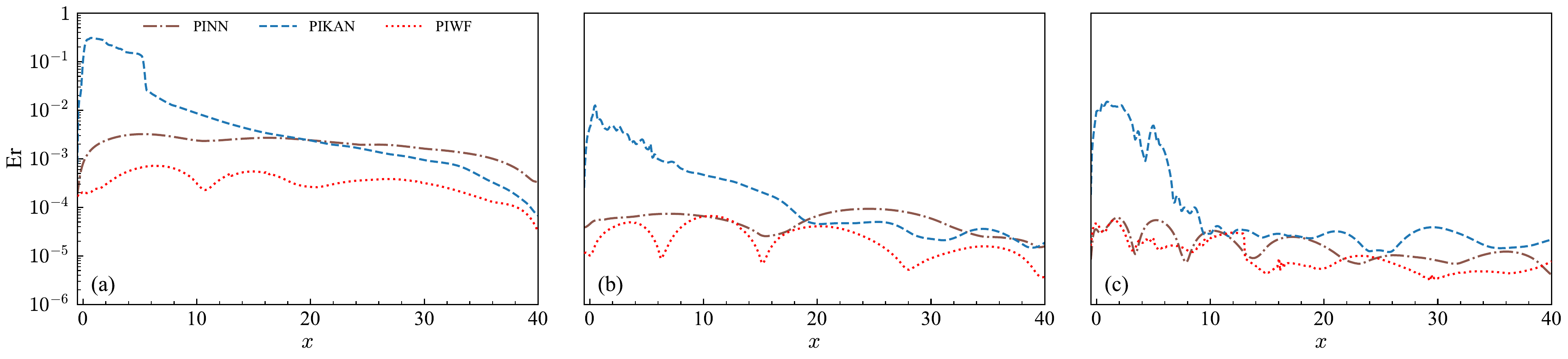}
		\caption{Relative error profiles for Kovasznay flow at ${\rm Re}=1000$: (a) streamwise velocity $u$, (b) transverse velocity $v$, and (c) pressure $p$. The profiles are averaged in the transverse direction.}
	\label{kovpreder}
\end{figure}

\begin{table}[htbp]
	\centering
	\footnotesize
	\setlength{\tabcolsep}{5pt}
		\caption{Mean relative errors for Kovasznay flow at ${\rm Re}=1000$.}
	\label{tab:kov_errors}
	\renewcommand{\arraystretch}{1.1}
		\begin{tabular}{lccc}
		\toprule
		Model & $u$ & $v$ & $p$ \\
		\midrule
			PINN  & $0.20\%$ & $0.0055\%$ & $0.0018\%$ \\
			PIKAN & $3.17\%$ & $0.073\%$ & $0.114\%$ \\
			PIWF  & $0.03\%$ & $0.0026\%$ & $0.0012\%$ \\
		\bottomrule
	\end{tabular}
\end{table}

The relative error profiles in Fig.~\ref*{kovpreder} and the mean relative errors in Table~\ref*{tab:kov_errors} further quantify the advantage of PIWF. For the streamwise velocity, PIWF reduces the mean relative error from $0.20\%$ for PINN and $3.17\%$ for PIKAN to $0.03\%$. The same trend is observed for the transverse velocity and pressure, although their absolute ranges are much smaller. This is important because the transverse velocity in Kovasznay flow is more than two orders of magnitude smaller than the dominant streamwise component; a model that mainly fits the largest-amplitude variable can therefore achieve a visually reasonable $u$ field while still missing the weaker coupled response. PIWF reduces this imbalance and gives a more consistent reconstruction of all three variables.

\begin{figure}[htbp]
	\centering
	\includegraphics[width=0.9\textwidth]{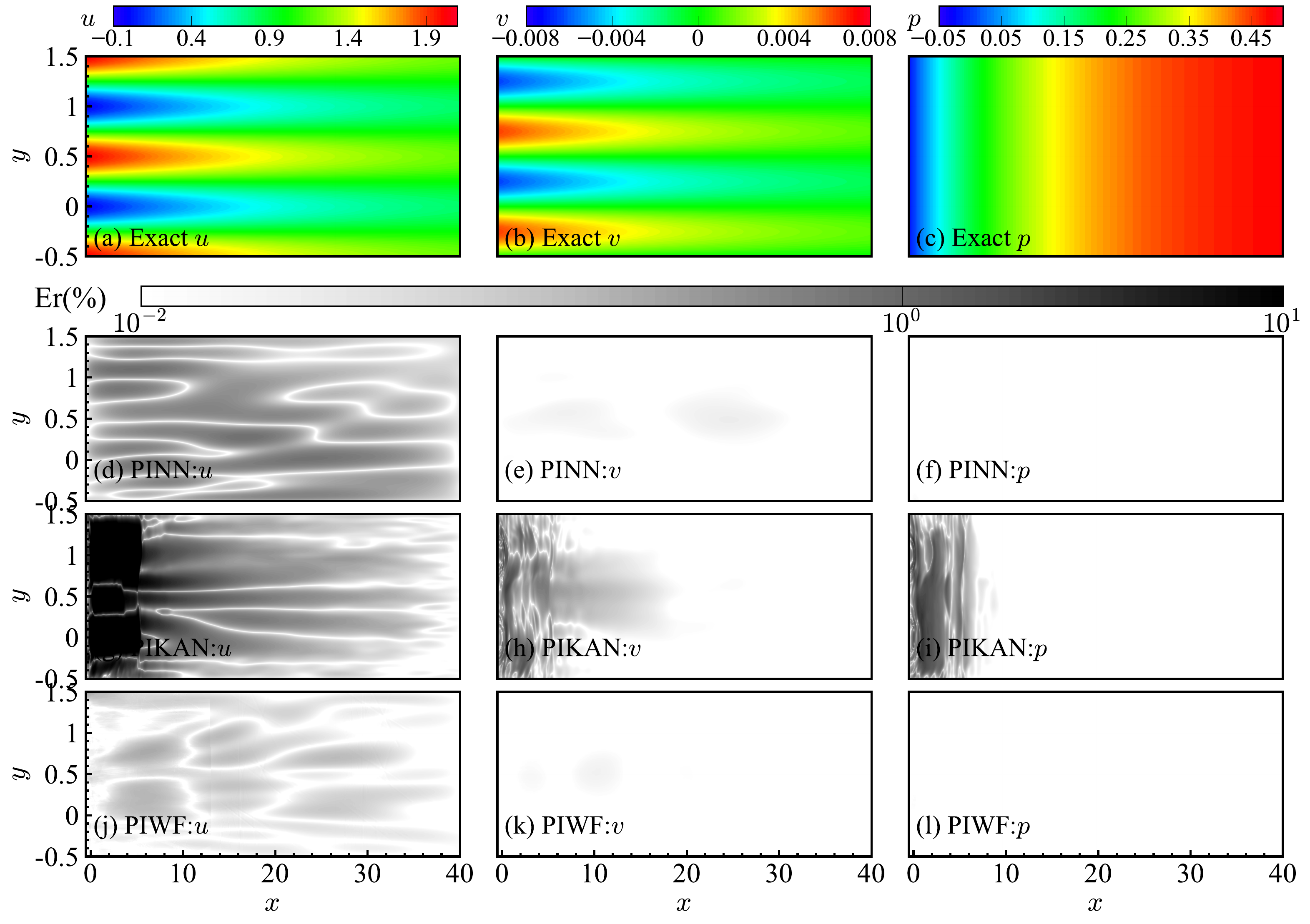}
			\caption{Reference fields and relative errors of PINN, PIKAN, and PIWF for Kovasznay flow at ${\rm Re}=1000$. From top to bottom: analytical solution and relative errors of PINN, PIKAN, and PIWF. From left to right: $u$, $v$, and $p$.}
	\label{kovcontour}
\end{figure}

The contour comparison in Fig.~\ref*{kovcontour} confirms that the error reduction is spatially coherent rather than confined to a small set of sample points. PINN produces larger errors in the near-wake region and along portions of the downstream decay, while PIKAN shows broader error bands across all three variables. PIWF gives the weakest and most localized relative-error contours, particularly in the streamwise velocity and pressure fields, where the exponential decay controls the downstream wake structure. These results show that the wavelet--Fourier representation is also effective for smooth viscous wakes, not only for sharp fronts or unsteady vortex shedding.

\subsection{Taylor--Green vortex flow}\label{sec:taylor_green}

The Taylor--Green vortex is then considered to examine the reconstruction of unsteady vortical motion in a periodic domain. In contrast to the Kovasznay benchmark, the flow evolves in time and contains interacting vortical structures whose amplitudes decay under viscosity. This case is used to evaluate whether the model can recover a two-dimensional incompressible Navier--Stokes solution from initial and boundary data while using only residual points in the interior. The governing equations are
\begin{equation}
	\begin{aligned}
		\frac{\partial u}{\partial x}+\frac{\partial v}{\partial y} &= 0,\\
		\frac{\partial u}{\partial t}+u\frac{\partial u}{\partial x}+v\frac{\partial u}{\partial y}
		&= -\frac{\partial p}{\partial x}
		+\frac{1}{\rm Re}\left(\frac{\partial^2 u}{\partial x^2}+\frac{\partial^2 u}{\partial y^2}\right),\\
		\frac{\partial v}{\partial t}+u\frac{\partial v}{\partial x}+v\frac{\partial v}{\partial y}
		&= -\frac{\partial p}{\partial y}
		+\frac{1}{\rm Re}\left(\frac{\partial^2 v}{\partial x^2}+\frac{\partial^2 v}{\partial y^2}\right).
	\end{aligned}
	\label{eq:tg-ns}
\end{equation}
The reference solution is generated in vorticity form,
\begin{equation}
	\frac{\partial \omega}{\partial t}
	+u\frac{\partial \omega}{\partial x}
	+v\frac{\partial \omega}{\partial y}
	=\frac{1}{\rm Re}
	\left(\frac{\partial^2\omega}{\partial x^2}
	+\frac{\partial^2\omega}{\partial y^2}\right),
	\qquad
	\omega=\frac{\partial v}{\partial x}-\frac{\partial u}{\partial y}.
	\label{eq:tg-vorticity}
\end{equation}
The Reynolds number is set to ${\rm Re}=1000$, and the periodic computational domain is $(x,y)\in[0,2\pi)\times[0,2\pi)$ over $t\in[0,30]$. The initial vorticity is prescribed as
\begin{equation}
	\begin{aligned}
		x_\theta &= x\cos\theta-y\sin\theta,\\
		y_\theta &= x\sin\theta+y\cos\theta,\\
		\omega(x,y,0) &= 2\cos(2x_\theta)\cos(2y_\theta)+\kappa,
	\end{aligned}
	\label{eq:tg-initial}
\end{equation}
where $\theta$ is uniformly sampled from $(-10^\circ,10^\circ)$ and $\kappa$ is sampled from a zero-mean normal distribution with variance $10^{-3}$. The reference field is first computed on a $256\times256$ grid and then downsampled by a factor of four to obtain the $64\times64$ fields used for evaluation. During training, high-fidelity reference data are supplied only on the initial plane and on the periodic boundaries; the remaining interior samples enter the loss only through the residual of Eq.~\ref{eq:tg-ns}. The PINN, PIKAN, and PIWF models use the same input vector $(x,y,t)$, the same output vector $(u,v,p)$, and the same depth and width as in the cylinder-wake case, namely four hidden layers with 128 neurons per layer. All models are trained for 200000 epochs with the Adam optimizer at a learning rate of 0.001.

\begin{figure}[htbp]
	\centering
	\includegraphics[width=0.9\textwidth]{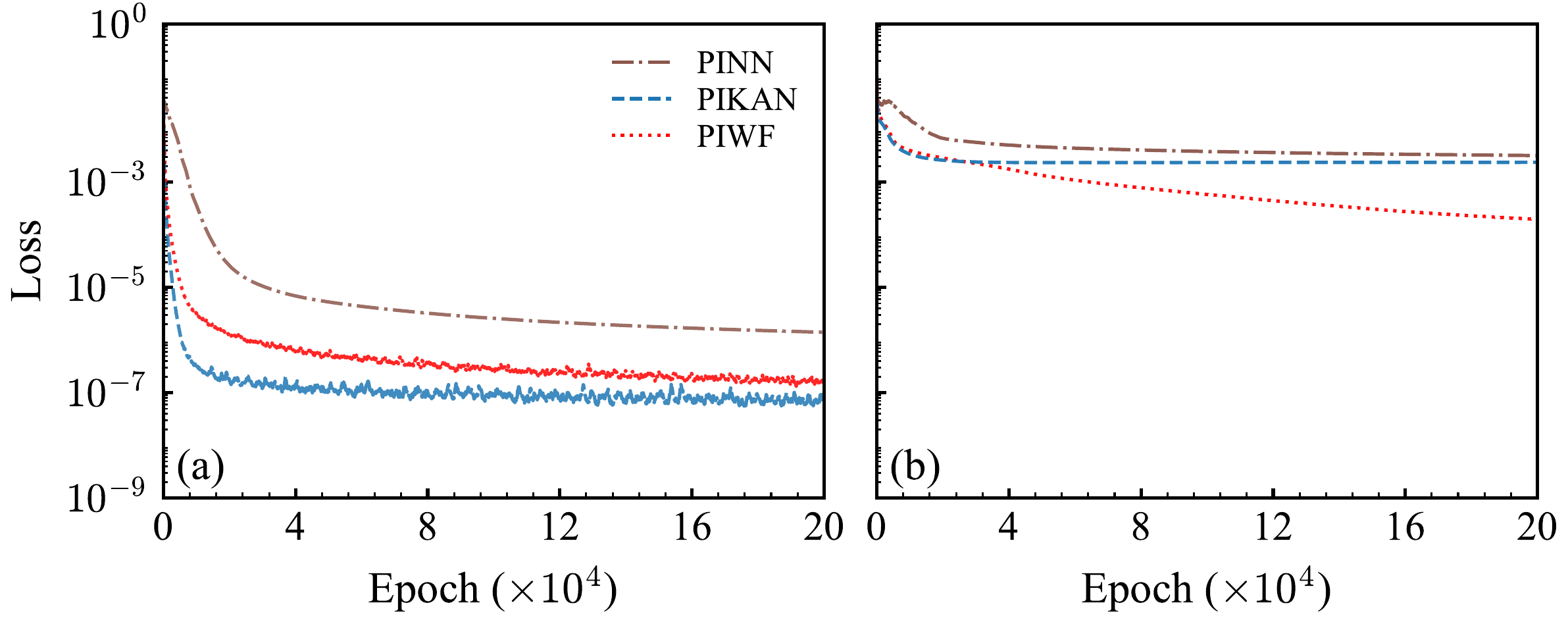}
			\caption{Learning curves of the PINN, PIKAN, and PIWF models for the Taylor--Green vortex flow at ${\rm Re}=1000$: (a) training loss; (b) testing loss.}
	\label{tgloss}
\end{figure}

The learning curves in Fig.~\ref{tgloss} show that all three models reduce the physics-informed objective, but their testing errors separate once the vortical field becomes more difficult to reconstruct from sparse constraints. PIWF maintains the lowest testing loss over the later training stage, indicating that the combined wavelet and Fourier-basis representation gives a more accurate reconstruction for this periodic vortical flow than the two baseline architectures considered here.

\begin{figure}[htbp]
	\centering
	\includegraphics[width=0.9\textwidth]{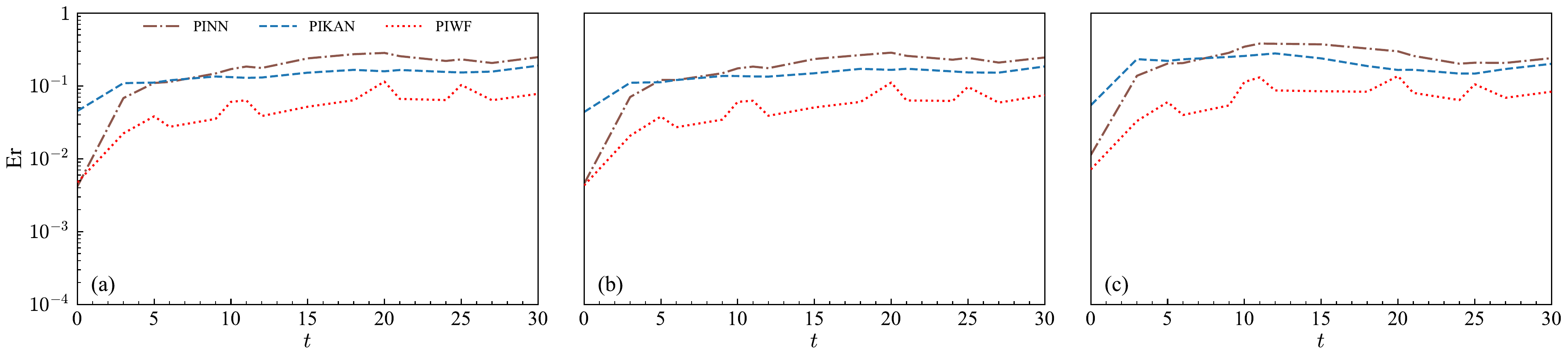}
		\caption{Temporal evolution of the relative $L^2$ errors for the Taylor--Green vortex flow at ${\rm Re}=1000$: (a) streamwise velocity $u$, (b) transverse velocity $v$, and (c) pressure $p$.}
	\label{tgpreder}
\end{figure}

Figure~\ref{tgpreder} compares the relative $L^2$ errors of the three predicted fields over the full time interval. The error curves show that PIWF gives lower errors for both velocity components and pressure over most of the interval. This comparison is particularly relevant for the pressure field because pressure is constrained indirectly through its gradients in the momentum equations, and pressure errors can affect the phase and amplitude of the predicted velocity field.

\begin{figure}[H]
	\centering
	\includegraphics[width=0.9\textwidth]{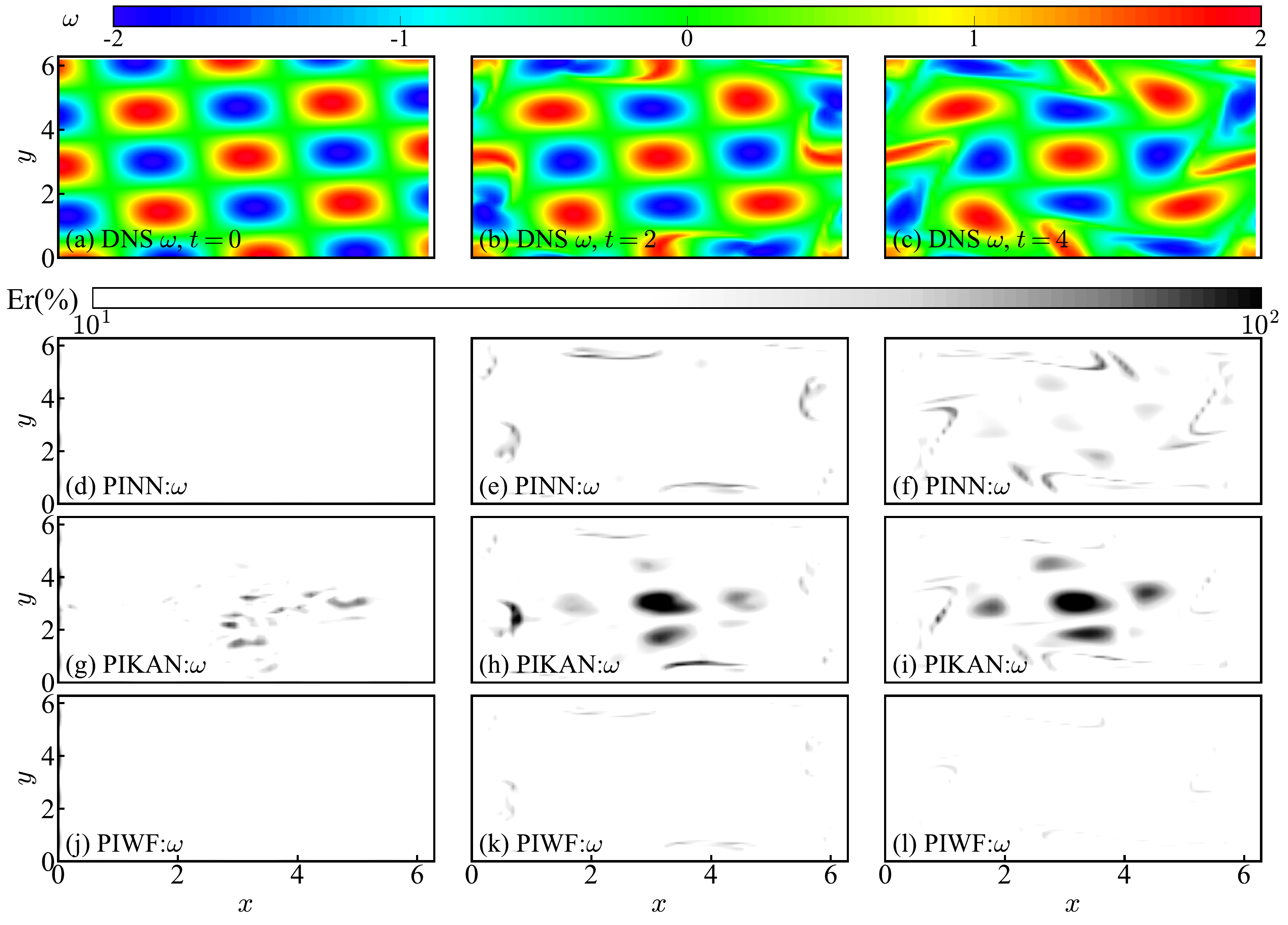}
			\caption{Reference vorticity and relative errors of PINN, PIKAN, and PIWF for the Taylor--Green vortex flow at $t=0$, 2, and 4.}
	\label{tgcontour1}
\end{figure}

\begin{figure}[H]
	\centering
	\includegraphics[width=0.9\textwidth]{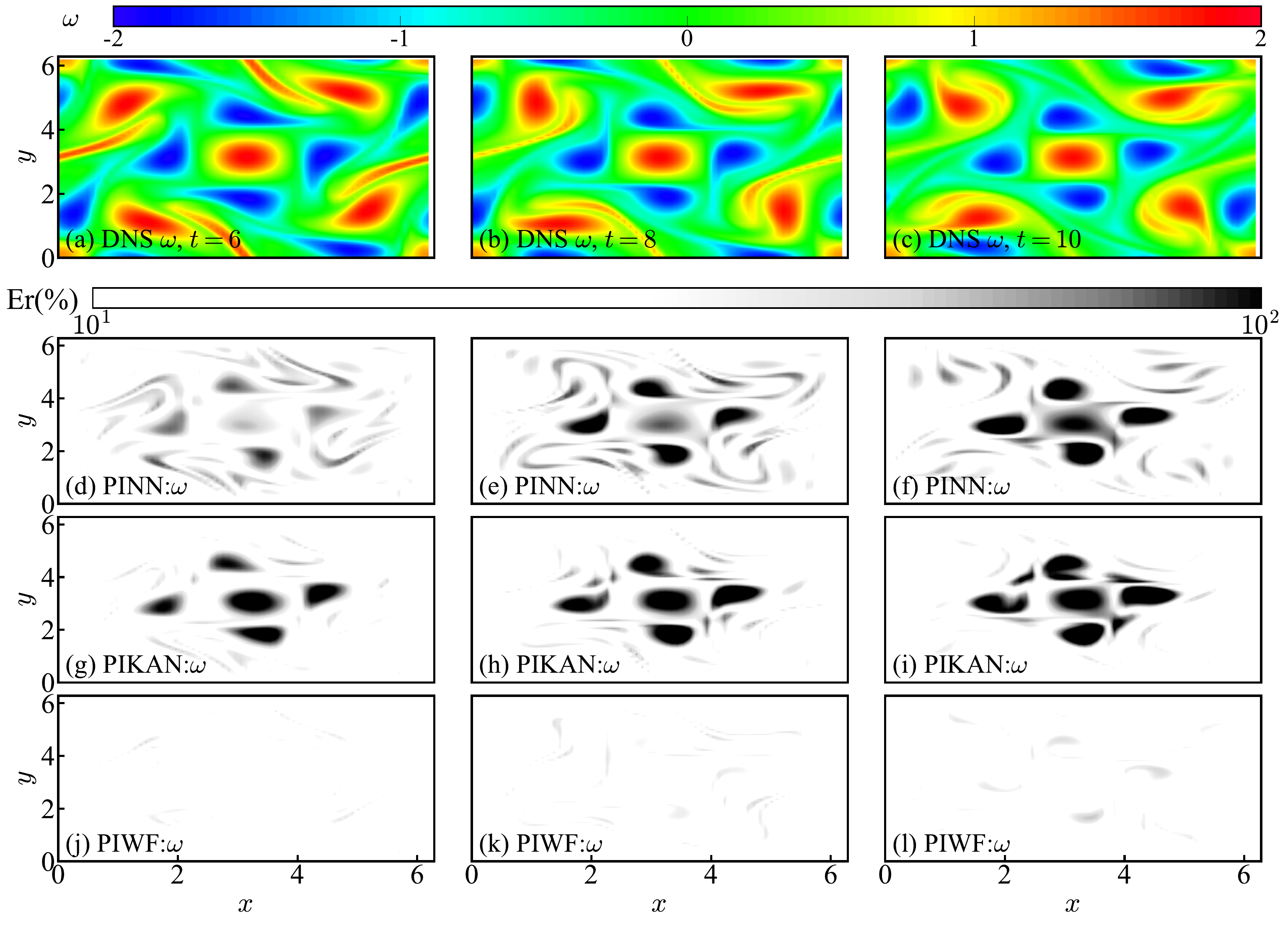}
			\caption{Reference vorticity and relative errors of PINN, PIKAN, and PIWF for the Taylor--Green vortex flow at $t=6$, 8, and 10.}
	\label{tgcontour2}
\end{figure}

The vorticity contours in Figs.~\ref{tgcontour1} and \ref{tgcontour2} provide a stricter spatial diagnostic than the primary velocity fields because vorticity depends on velocity gradients. At early times, all models recover the large-scale vortex arrangement, but the error fields already show stronger localization for PIWF. As the flow evolves, PINN and PIKAN develop broader error regions around the vortex cores and inter-vortex shear layers. PIWF gives smaller and more spatially confined errors over the selected time interval, which indicates improved preservation of both the periodic large-scale pattern and the local gradient information.

\subsection{Two-dimensional cylinder wake flow}

Flow past a circular cylinder is a classical benchmark problem in fluid mechanics because it contains boundary-layer separation, shear-layer roll-up, and periodic vortex shedding. It therefore provides a stringent test of whether a physics-informed model can preserve coherent wake dynamics and derivative-based flow diagnostics, rather than merely approximate the primary velocity and pressure fields. In this study, two-dimensional cylinder wake flow is used to assess the capability of neural-network-based models in capturing complex, time-dependent flow physics. The cylinder wake flow is governed by the incompressible Navier--Stokes equations, namely
\begin{equation}
	\begin{aligned}
		\frac{\partial u}{\partial x} + \frac{\partial v}{\partial y} &= 0, \\
		\frac{\partial u}{\partial t} + u \frac{\partial u}{\partial x} + v \frac{\partial u}{\partial y} 
		&= -\frac{\partial p}{\partial x} + \frac{1}{\rm Re}\left( \frac{\partial^{2} u}{\partial x^{2}} + \frac{\partial^{2} u}{\partial y^{2}} \right), \\
		\frac{\partial v}{\partial t} + u \frac{\partial v}{\partial x} + v \frac{\partial v}{\partial y} 
		&= -\frac{\partial p}{\partial y} + \frac{1}{\rm Re}\left( \frac{\partial^{2} v}{\partial x^{2}} + \frac{\partial^{2} v}{\partial y^{2}} \right).
	\end{aligned}
	\label{eq:ns-wake}
\end{equation}
Here, $u$ and $v$ denote the velocity components in the $x$- and $y$- directions, respectively, and $p$ is the modified pressure divided by the constant density. The Reynolds number is set to $\rm Re=100$. 
The reference solution is generated using the spectral/$hp$ element solver Nektar~\cite{raissi2019physics}. The boundary conditions are as follows: a uniform free-stream velocity is imposed at the inlet ($x=-15$), a zero-pressure outflow condition is applied at the outlet ($x=25$), and periodic boundary conditions are enforced along the transverse boundaries. The global computational domain is $(x, y) \in [-15, 25] \times [-8, 8]$. In the present study, we focus the analysis on the wake region, specifically the subdomain $[1, 8] \times [-2, 2]$ over the time interval $t \in [0, 7]$.

\begin{figure}[htbp] 
	\centering
	\includegraphics[width=0.9\textwidth]{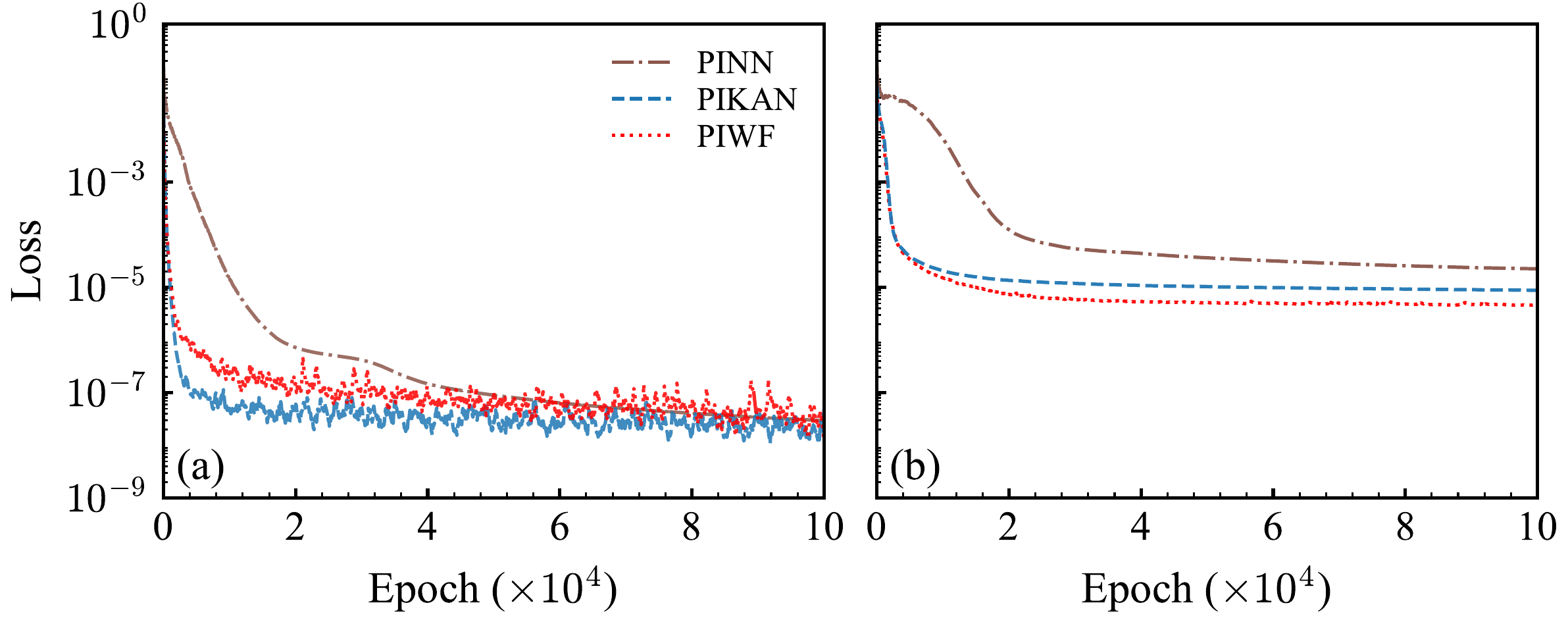}
		\caption{Learning curves of the PINN, PIKAN, and PIWF models for the two-dimensional cylinder wake flow at $\rm Re=100$: (a) training loss; (b) testing loss.}
	\label{cypredloss}
\end{figure}

Training data are generated as follows: 100 points are sampled along the transverse boundaries and 50 points along the inlet and outlet boundaries using Latin hypercube sampling (LHS) strategy to enforce the Dirichlet conditions. Within the spatio-temporal domain, 140,000 collocation points are uniformly randomly sampled to evaluate the physics-informed residual in  Eq. \ref{eq15}.
An additional 8000 points are reserved as a test set for quantitative evaluation.

All neural-network models (PINN, PIKAN and PIWF) consist of 4 hidden layers with 128 neurons per layer. The input variables comprise the spatio-temporal coordinates $(x,y,t)$, while the output vector consists of the velocity components $(u, v)$ and pressure $p$. Thus, both the input and output layers contain 3 neurons. All models are trained for 100000 epochs to ensure that the loss function converges to a stable minimum.

\begin{figure}[htbp]
	\centering
	\includegraphics[width=0.8\textwidth]{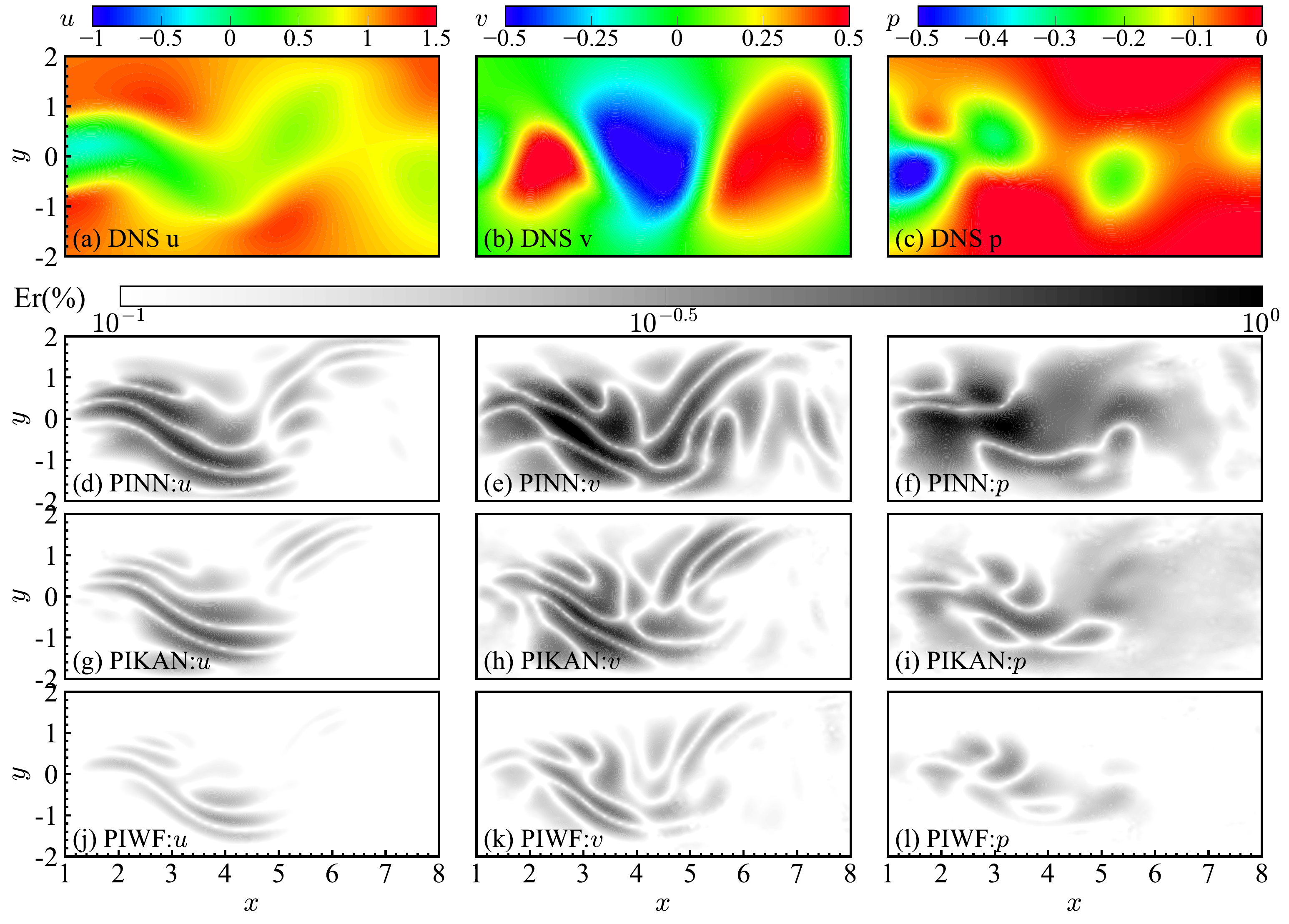}
	\caption{Relative error contours of the velocity components $(u,v)$ and pressure $(p)$ for the cylinder wake flow at $t=1$. From top to bottom: DNS, PINN, PIKAN, and PIWF results. From left to right: $u$, $v$, and $p$.}
	\label{cylinderres} 
\end{figure}

Fig. \ref{cypredloss} illustrates the learning curves of the PINN, PIKAN, and PIWF models for the two-dimensional cylinder wake flow at $\rm Re=100$. During the later training stages, all three models reach similar training loss values, indicating that the imposed boundary and residual constraints are comparably satisfied. The testing loss, however, reveals clear differences in the recovered wake physics. PIWF consistently achieves the lowest testing error, suggesting that the hybrid wavelet--Fourier representation generalizes better to interior wake structures that are not directly prescribed by the boundary data.

\begin{figure}[htbp]
	\centering
	\includegraphics[width=0.8\textwidth]{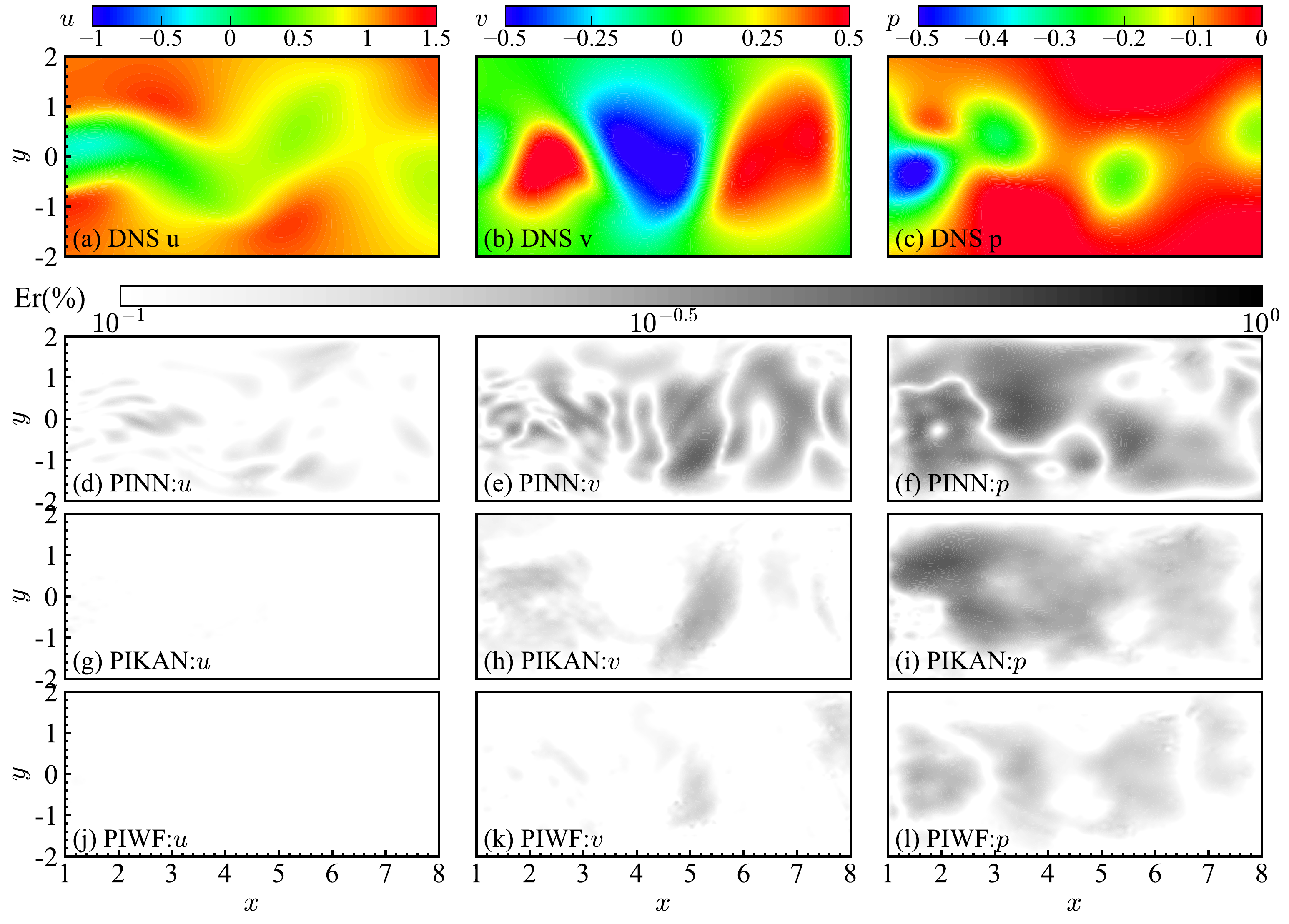}
	\caption{Relative error contours of the velocity components $(u,v)$ and pressure $(p)$ for the cylinder wake flow at $t=7$. From top to bottom: DNS, PINN, PIKAN, and PIWF results. From left to right: $u$, $v$, and $p$.}
	\label{cylinderres2} 
\end{figure}

The relative error contours in Figs. \ref{cylinderres} and \ref{cylinderres2} further show that PIWF yields lower and more spatially localized errors than the baseline models. The improvement is most apparent in regions associated with separated shear layers and wake vortices, where the gradients of velocity and pressure are large. Accurately resolving these regions is essential because errors in the transverse velocity $v$ and pressure $p$ directly affect the phase, strength, and induced pressure field of the vortex street. PIWF reduces the disparity among the errors of $u$, $v$, and $p$, whereas PINN and PIKAN exhibit stronger localized deviations in the $v$ and $p$ fields.

\begin{figure}[htbp] 
	\centering
	\includegraphics[width=0.9\textwidth]{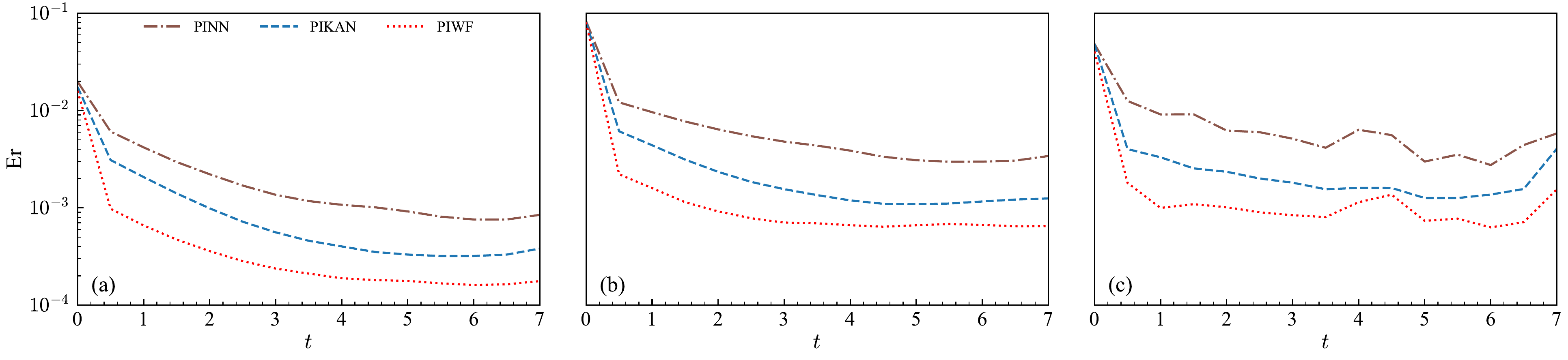}
	\caption{Temporal evolution of the relative $L^2$ errors for the cylinder wake flow at $\rm Re=100$: (a) streamwise velocity $u$, (b) transverse velocity $v$, (c) pressure $p$.}
	\label{cypreder}
\end{figure}

The temporal evolution of the relative $L^2$ errors, displayed in Fig. \ref{cypreder}, provides a clear quantitative comparison of the predictive performance across these three models. A consistent hierarchy is observed: PIKAN outperforms the standard PINN, while the proposed PIWF achieves the lowest error profiles throughout the entire temporal domain. These results are further demonstrated in Table \ref{tab:cylinder_L2_errors}, which reports the relative $L^2$ errors for the velocity components and pressure at discrete time instants as well as the time-averaged values. Notably, PIWF demonstrates a sustained and substantial advantage across all investigated temporal instances.

\begin{table}
	\centering
	\footnotesize
	\setlength{\tabcolsep}{4pt}
	\caption{Relative $L^2$ errors for the two-dimensional cylinder wake flow at $\rm Re=100$.}
	\label{tab:cylinder_L2_errors}
	\renewcommand{\arraystretch}{1.1}
	\resizebox{\textwidth}{!}{%
	\begin{tabular}{lcccccccc}
		\toprule
		Model / Time & 1 & 2 & 3 & 4 & 5 & 6 & 7 & Average \\
		\midrule
		$u$ (PINN)  & $6.074 \times 10^{-3}$ & $4.178 \times 10^{-3}$ & $2.956 \times 10^{-3}$ & $2.212 \times 10^{-3}$ & $1.693 \times 10^{-3}$ & $1.360 \times 10^{-3}$ & $1.172 \times 10^{-3}$ & $2.806 \times 10^{-3}$ \\
		$u$ (PIKAN) & $3.095 \times 10^{-3}$ & $2.069 \times 10^{-3}$ & $1.406 \times 10^{-3}$ & $9.868 \times 10^{-4}$ & $7.194 \times 10^{-4}$ & $5.591 \times 10^{-4}$ & $4.583 \times 10^{-4}$ & $1.328 \times 10^{-3}$ \\
		$u$ (PIWF)  & $9.769 \times 10^{-4}$ & $6.584 \times 10^{-4}$ & $4.716 \times 10^{-4}$ & $3.589 \times 10^{-4}$ & $2.830 \times 10^{-4}$ & $2.373 \times 10^{-4}$ & $2.106 \times 10^{-4}$ & $4.567 \times 10^{-4}$ \\
		\midrule
		$v$ (PINN)  & $1.212 \times 10^{-2}$ & $9.613 \times 10^{-3}$ & $7.710 \times 10^{-3}$ & $6.396 \times 10^{-3}$ & $5.447 \times 10^{-3}$ & $4.790 \times 10^{-3}$ & $4.354 \times 10^{-3}$ & $7.204 \times 10^{-3}$ \\
		$v$ (PIKAN) & $6.117 \times 10^{-3}$ & $4.395 \times 10^{-3}$ & $3.127 \times 10^{-3}$ & $2.342 \times 10^{-3}$ & $1.846 \times 10^{-3}$ & $1.554 \times 10^{-3}$ & $1.350 \times 10^{-3}$ & $2.962 \times 10^{-3}$ \\
		$v$ (PIWF)  & $2.212 \times 10^{-3}$ & $1.593 \times 10^{-3}$ & $1.141 \times 10^{-3}$ & $9.184 \times 10^{-4}$ & $7.809 \times 10^{-4}$ & $7.054 \times 10^{-4}$ & $6.929 \times 10^{-4}$ & $1.149 \times 10^{-3}$ \\
		\midrule
		$p$ (PINN)  & $1.254 \times 10^{-2}$ & $9.112 \times 10^{-3}$ & $9.144 \times 10^{-3}$ & $6.225 \times 10^{-3}$ & $5.983 \times 10^{-3}$ & $5.134 \times 10^{-3}$ & $4.141 \times 10^{-3}$ & $7.469 \times 10^{-3}$ \\
		$p$ (PIKAN) & $4.014 \times 10^{-3}$ & $3.305 \times 10^{-3}$ & $2.541 \times 10^{-3}$ & $2.346 \times 10^{-3}$ & $2.000 \times 10^{-3}$ & $1.812 \times 10^{-3}$ & $1.550 \times 10^{-3}$ & $2.510 \times 10^{-3}$ \\
		$p$ (PIWF)  & $1.802 \times 10^{-3}$ & $9.964 \times 10^{-4}$ & $1.090 \times 10^{-3}$ & $1.013 \times 10^{-3}$ & $8.990 \times 10^{-4}$ & $8.393 \times 10^{-4}$ & $8.027 \times 10^{-4}$ & $1.063 \times 10^{-3}$ \\
		\bottomrule
	\end{tabular}%
	}
\end{table}

To further evaluate the models' capability in resolving fine-scale vortical structures, the vorticity $\omega_{z}$ and the $Q$-criterion are computed from the predicted velocity fields using second-order central difference scheme. These quantities are defined by
\begin{equation}
	\omega_z = \frac{\partial v}{\partial x} - \frac{\partial u}{\partial y} \quad {\rm{and}} \quad\ Q = \frac{1}{2} \left( \|\boldsymbol{\Omega}\|^2 - \|\mathbf{S}\|^2 \right),
	\label{omega}
\end{equation}
where $\bm{\Omega}$ is the rotation-rate tensor, which is the antisymmetric part of the velocity gradient tensor. $\|\boldsymbol{\Omega}\|^2$ is defined by
\begin{equation}
	\|\boldsymbol{\Omega}\|^2 = \frac{1}{2} \left( \frac{\partial v}{\partial x} - \frac{\partial u}{\partial y} \right)^2.
\end{equation}
The tensor $\mathbf{S}$ is the strain-rate tensor, which is the symmetric part of the velocity gradient tensor. $\|\mathbf{S}\|^2$ is defined by

\begin{equation}
	\|\mathbf{S}\|^2 = \left( \frac{\partial u}{\partial x} \right)^2 + \left( \frac{\partial v}{\partial y} \right)^2 + \frac{1}{2} \left( \frac{\partial u}{\partial y} + \frac{\partial v}{\partial x} \right)^2.
\end{equation}

\begin{figure}[htbp]
	\centering
	\includegraphics[width=0.9\textwidth]{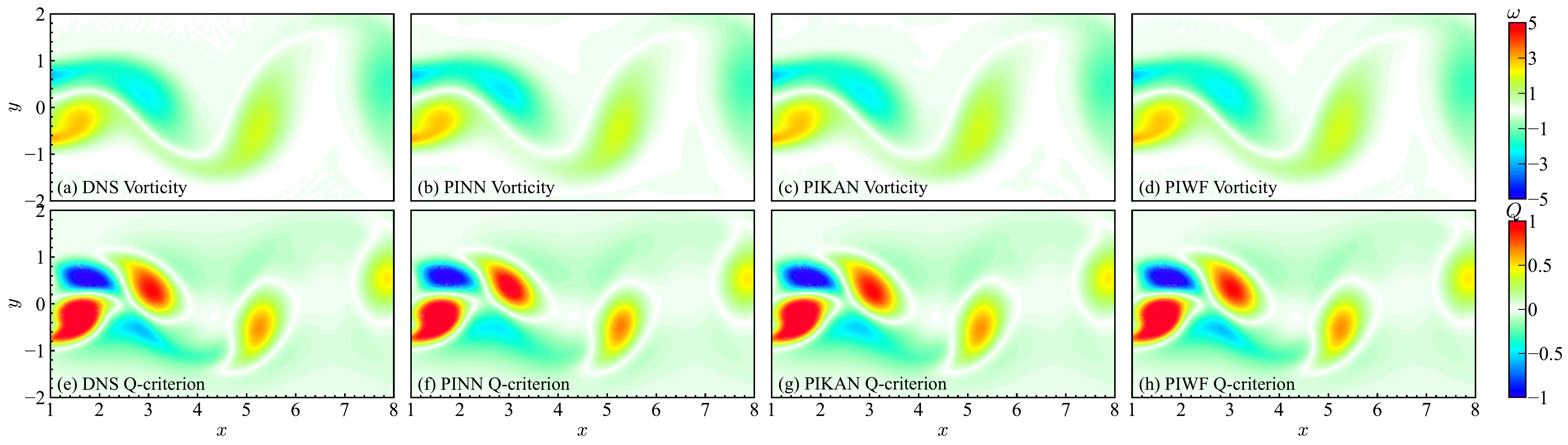}
	\caption{Contours of vorticity ($\omega_{z}$) and the $Q$-criterion for the cylinder wake flow predicted by the DNS, PINN, PIKAN, and PIWF models. Top row: vorticity; bottom row: $Q$. From left to right: DNS, PINN, PIKAN, and PIWF. }
	\label{cylinderpred} 
\end{figure}

\begin{table}
	\centering
	\footnotesize
	\setlength{\tabcolsep}{4pt}
	\caption{Errors for vorticity ($\omega_z$) and the $Q$-criterion in the two-dimensional cylinder wake flow at $\rm Re=100$. Vorticity errors are absolute errors, while $Q$ errors are relative errors.}
	\label{tab:vorticity_q_errors}
	\renewcommand{\arraystretch}{1.1}
	\resizebox{\textwidth}{!}{%
	\begin{tabular}{lcccccccc}
		\toprule
		Model / Time & 1 & 2 & 3 & 4 & 5 & 6 & 7 & Average \\
		\midrule
		$\omega_z$ (PINN)  & $2.929 \times 10^{-2}$ & $1.654 \times 10^{-2}$ & $9.834 \times 10^{-3}$ & $6.539 \times 10^{-3}$ & $5.277 \times 10^{-3}$ & $4.855 \times 10^{-3}$ & $5.632 \times 10^{-3}$ & $1.114 \times 10^{-2}$ \\
		$\omega_z$ (PIKAN) & $1.507 \times 10^{-2}$ & $7.293 \times 10^{-3}$ & $3.632 \times 10^{-3}$ & $2.074 \times 10^{-3}$ & $1.579 \times 10^{-3}$ & $1.381 \times 10^{-3}$ & $1.447 \times 10^{-3}$ & $4.639 \times 10^{-3}$ \\
		$\omega_z$ (PIWF)  & $5.237 \times 10^{-3}$ & $2.602 \times 10^{-3}$ & $1.510 \times 10^{-3}$ & $1.143 \times 10^{-3}$ & $1.053 \times 10^{-3}$ & $1.095 \times 10^{-3}$ & $1.113 \times 10^{-3}$ & $1.965 \times 10^{-3}$ \\
		\midrule
		$Q$ (PINN)  & $1.506 \times 10^{-1}$ & $1.037 \times 10^{0}$ & $8.984 \times 10^{-2}$ & $8.134 \times 10^{-2}$ & $7.616 \times 10^{-2}$ & $7.524 \times 10^{-2}$ & $6.221 \times 10^{-2}$ & $2.247 \times 10^{-1}$ \\
		$Q$ (PIKAN) & $7.248 \times 10^{-2}$ & $1.081 \times 10^{-1}$ & $3.280 \times 10^{-2}$ & $3.417 \times 10^{-2}$ & $1.488 \times 10^{-2}$ & $1.761 \times 10^{-2}$ & $1.607 \times 10^{-2}$ & $4.230 \times 10^{-2}$ \\
		$Q$ (PIWF)  & $3.066 \times 10^{-2}$ & $1.475 \times 10^{-1}$ & $2.356 \times 10^{-2}$ & $2.368 \times 10^{-2}$ & $1.049 \times 10^{-2}$ & $1.445 \times 10^{-2}$ & $1.123 \times 10^{-2}$ & $3.737 \times 10^{-2}$ \\
		\bottomrule
	\end{tabular}%
	}
\end{table}

\begin{figure}[htbp]
	\centering
	\includegraphics[width=0.9\textwidth]{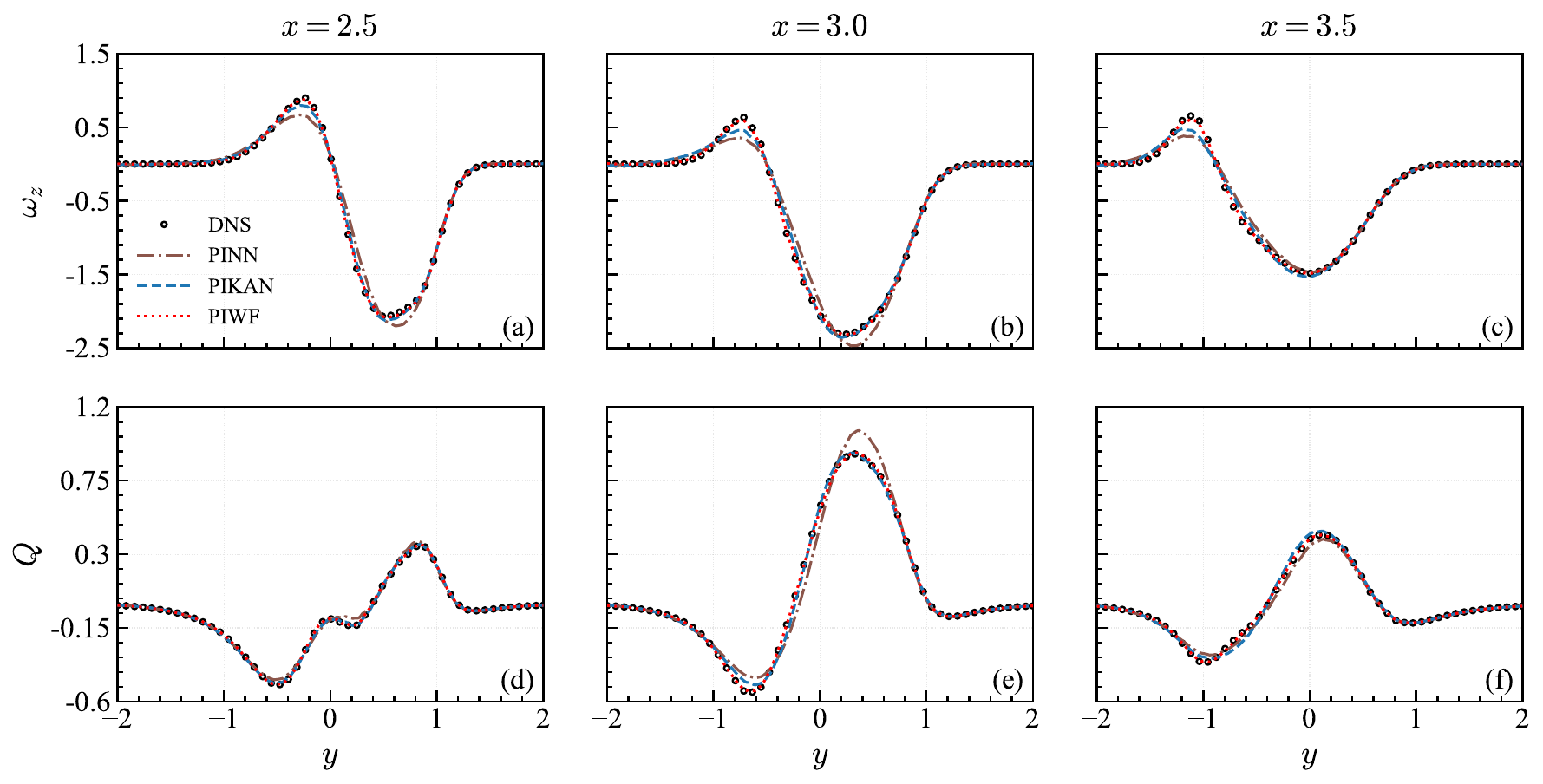}
	\caption{Profiles of vorticity ($\omega_{z}$) and the $Q$-criterion along the lines $x=2.5$, $3.0$, and $3.5$ at $t=1$ predicted by the PINN, PIKAN, and PIWF models. Top row: vorticity; bottom row: $Q$.}
	\label{cypred}
\end{figure}

The resulting contours and profiles of vorticity ($\omega_{z}$) and $Q$ are displayed in Figs. \ref{cylinderpred} and \ref{cypred}, respectively. Compared with the primary velocity and pressure fields, these derivative-based quantities provide a more stringent assessment of whether the predicted wake contains the correct rotational and strain-dominated structures. Both PINN and PIKAN deviate near local extrema in vorticity and $Q$, with the largest deviations observed for PINN. Table \ref{tab:vorticity_q_errors} summarizes the absolute errors in vorticity and relative errors in $Q$ at selected temporal instants, together with the time-averaged errors. The relative error of $Q$ for PINN reaches order unity at some instants, indicating a substantial distortion of the balance between rotation and strain. PIWF yields derivative-based diagnostics that are more consistent with the reference fields, demonstrating that the multiresolution wavelet component helps preserve fine-scale vortical features in the wake.

\begin{figure}[htbp]
	\centering
	\includegraphics[width=0.8\textwidth]{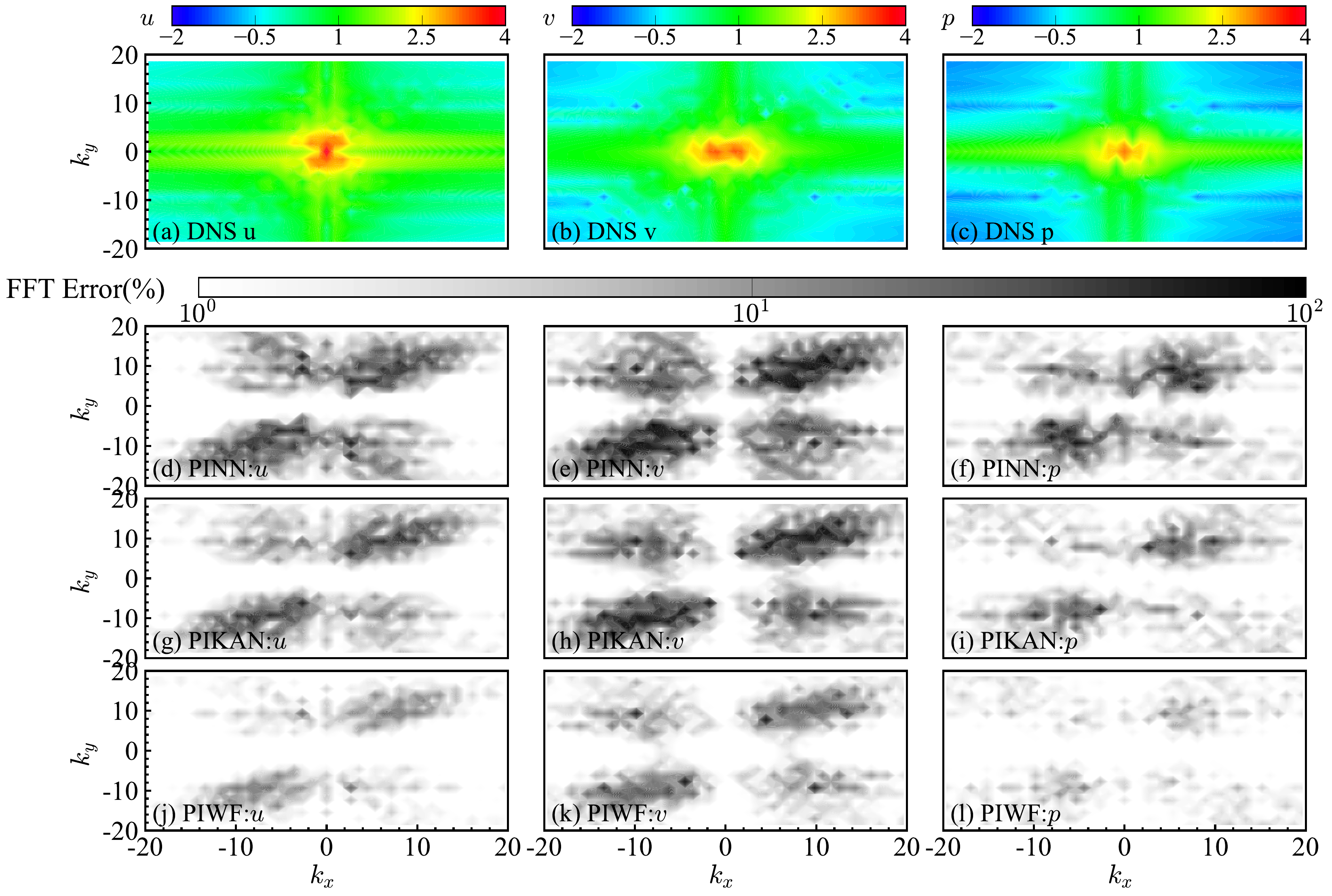}
	\caption{Relative error contours in wavenumber space for the cylinder wake flow at $t=1$. From top to bottom: DNS, PINN, PIKAN, and PIWF results. From left to right: streamwise velocity $u$, transverse velocity $v$, and pressure $p$.}
	\label{cylinderres3} 
\end{figure}

\begin{figure}[htbp]
	\centering
	\includegraphics[width=0.8\textwidth]{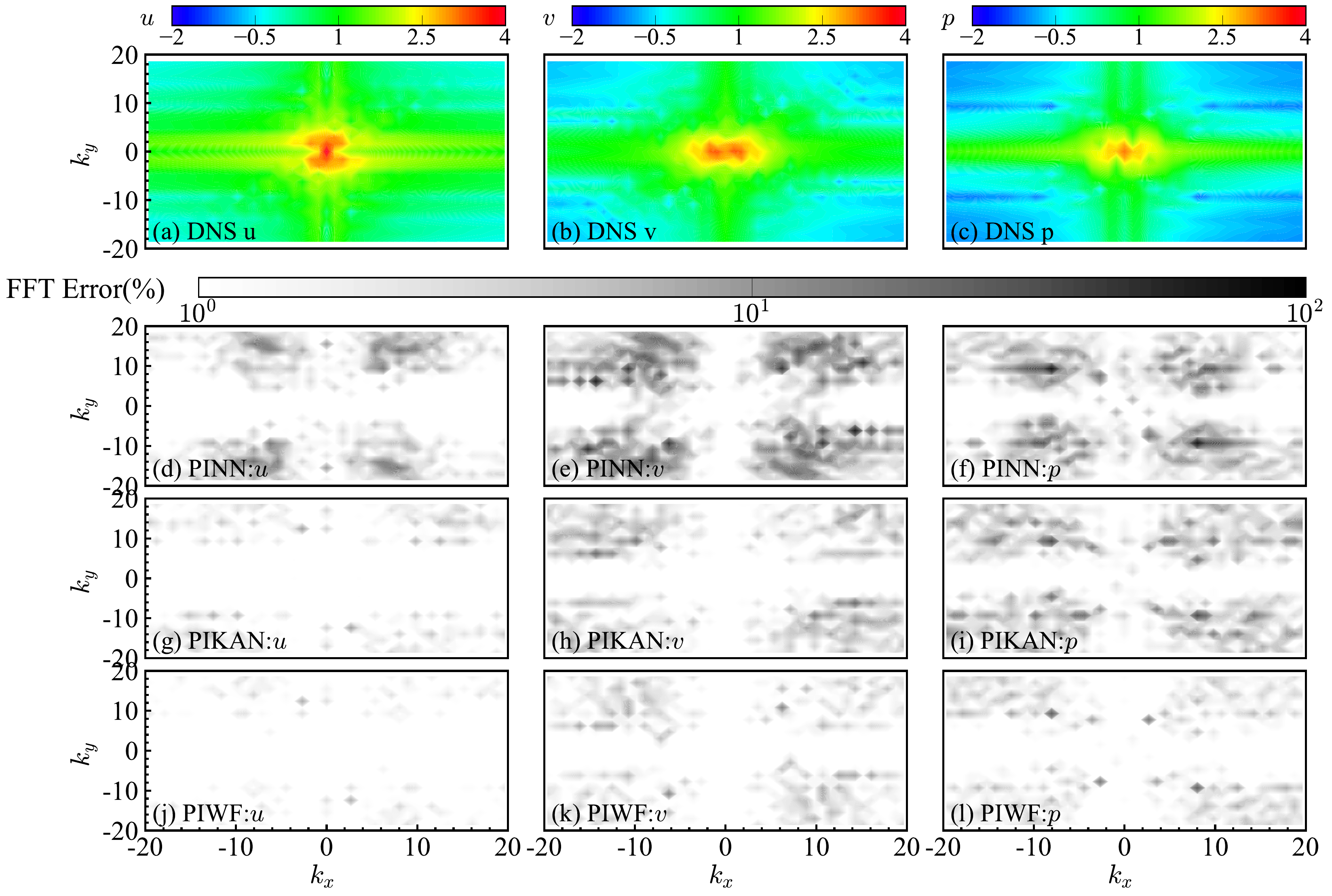}
	\caption{Relative error contours in wavenumber space for the cylinder wake flow at $t=7$. From top to bottom: DNS, PINN, PIKAN, and PIWF results. From left to right: streamwise velocity $u$, transverse velocity $v$, and pressure $p$.}
	\label{cylinderres4} 
\end{figure}

To further evaluate the models' ability to preserve the broadband energy distribution in the cylinder wake, we perform a two-dimensional fast Fourier transform (FFT) analysis in wavenumber space ($k_x$, $k_y$). The resulting spectra for the streamwise velocity $u$, transverse velocity $v$, and pressure $p$ at representative time instants $t=1$ and 7 are shown in Figs \ref{cylinderres3} and \ref{cylinderres4}. The top row presents the reference DNS spectra, while the lower rows display the relative FFT error (in percent, logarithmic scale) for the PINN, PIKAN, and PIWF predictions. In wavenumber space, both PINN and PIKAN exhibit increased spectral errors at moderate-to-high wavenumbers. These errors indicate reduced fidelity in capturing the fine-scale turbulent structures and the correct energy cascade, with energy leakage and spurious high-frequency components.

In contrast, PIWF gives relatively low FFT errors across the wavenumber spectrum and remains close to the DNS reference even at the smallest resolved scales. This indicates that PIWF accurately captures the coherent structures in the cylinder wake while introducing less spectral distortion than the baseline models. The wavelet layer improves the localized high-wavenumber content without introducing strong Gibbs-like oscillations, whereas the Fourier branch preserves the low-wavenumber coherent modes. Channel attention then adaptively weights these complementary representations, helping retain the broadband energy distribution of the wake.


Collectively, the spatial error distributions, derivative-based diagnostics, and spectral analyses show that PIWF improves point-wise reconstruction accuracy while better retaining the multiscale structures of the wake. This capability is particularly relevant for unsteady nonlinear wake dynamics, where accurate prediction requires simultaneous preservation of vortex position, vorticity magnitude, pressure response, and broadband spectral content. These results support the use of PIWF as a physics-informed representation for multiscale fluid problems. An ablation study and parameter sensitivity analysis are provided in Appendix~\ref{sec:ablation}.

\section{\label{sec:level1/6} Summary}
In this work, we propose a physics-informed wavelet-Fourier (PIWF) representation for resolving multiscale structures in fluid-dynamics problems. The central element is a dedicated wavelet layer that reformulates each network mapping as a wavelet-series approximation, enabling multilevel extraction of both global structures and localized sharp features such as discontinuities, steep gradients, and singularities. The PIWF representation comprises three parallel branches operating at complementary scales: a Fourier-basis layer for capturing global low-frequency components, the proposed wavelet layer for targeted extraction of high-frequency localized content, and a MLP layer for residual compensation. These heterogeneous outputs are dynamically fused through a channel attention mechanism that adaptively balances global spectral consistency with local feature resolution. By embedding the governing physical laws into the loss function, PIWF can compute forward solutions without high-fidelity interior reference data.

The numerical results show that this scale-aware representation improves the recovery of several physically important flow features. For Burgers' equation, PIWF more accurately localizes the shock-like velocity layer, especially at higher Reynolds number. For the shallow-water equations, it better preserves the wet--dry-front position and the coupled response of water depth and discharge. For Kovasznay flow at ${\rm Re}=1000$, PIWF gives a more accurate reconstruction of the long viscous wake and reduces the relative errors of the coupled velocity and pressure fields. For the Taylor--Green vortex at ${\rm Re}=1000$, PIWF better preserves the decaying vorticity field and reduces the physical-space errors. For the cylinder wake at $\rm Re=100$, PIWF improves the reconstruction of velocity, pressure, vorticity, and the $Q$-criterion, while also reducing spectral distortion across the wavenumber plane. These observations indicate that the advantage of PIWF is not limited to a lower global error metric; it lies in the improved preservation of the flow structures that control discontinuity propagation, free-surface-front dynamics, steady wake decay, vortex decay, and vortex shedding.

Overall, PIWF provides a scale-aware physics-informed representation for multiscale fluid problems involving sharp interfaces, vortex structures, and broadband energy distributions. Nevertheless, the current PIWF still has certain limitations. Its performance advantage is most pronounced in problems with relatively simple boundary conditions. For cases involving highly complex or irregular boundaries, the fixed global computational domain and uniform collocation strategy may lead to reduced efficiency and accuracy. Future work will focus on developing region-adaptive and block-wise training strategies that allow the wavelet layer to concentrate on challenging local regions while the Fourier-basis layer handles smoother areas, exploring boundary-aware collocation schemes for complex geometries, extending the formulation to three-dimensional transitional and turbulent flows, and incorporating adaptive loss weighting to further stabilize training. These developments would broaden the use of PIWF for physics-informed analysis of multiscale flow phenomena.

\section*{CRediT authorship contribution statement}\label{sec:level_CRediT}
{\bfseries Chao Wang}: Writing -- original draft, Funding acquisition, Resources, Supervision, Writing review \& editing;
{\bfseries Shilong Li}: Writing -- original draft, Validation,  Methodology, Formal analysis, Conceptualization; 
{\bfseries Yunpeng~Wang}: Writing review \& editing, Conceptualization, Supervision; {\bfseries Tianbai Xiao}: Writing review \& editing, Conceptualization, Supervision; 
{\bfseries Zelong Yuan}: Writing review \& editing, Methodology, Formal analysis, Conceptualization, Funding acquisition; 
{\bfseries Chenyue Xie}: Writing review \& editing, Methodology, Formal analysis, Conceptualization, Visualization; 
{\bfseries Chunyu Guo}: Writing review \& editing, Funding acquisition, Supervision.

\section*{Data availability}\label{sec:level_Data}
The source code, implementation details, example scripts, execution instructions, and all datasets required to reproduce the numerical experiments are provided in the GitHub repository at \url{https://github.com/OneMoreSpeed/PIWF}.
\section*{Declaration of competing interest}\label{sec:level_Declaration}
The authors declare that they have no known competing financial interests or personal relationships that could have appeared to influence the work reported in this paper.
\section*{Acknowledgment}\label{sec:level_Acknowledgment}
This work was supported by the National Natural Science Foundation of China (Grant Nos. 52425111, 52306193, 12388101, 12502261), the Natural Science Foundation of Hainan Province (No. 425QN376), the Natural Science Foundation of Shandong Province (No. ZR2024QA050), and the Basic Product Innovation Research Project (No. KY10100230067).

\appendix
\section{\label{sec:level3wavelt}Wavelet transform and wavelet series}

The wavelet transform is introduced to overcome the fixed window limitation of the short-time Fourier transform (STFT) by employing variable window functions that enable decomposition of a signal into components at different frequencies and scales. These window functions are generated from a mother wavelet ($\psi ( t )$) through scaling and translation operations. The mother wavelet should satisfy the admissibility and unit-energy conditions
\begin{equation}
	\begin{array} { l } { \int _ { - \infty } ^ { \infty } \psi ( t ) d t = 0 }\quad \textrm{and} \quad
		 \ { \int _ { - \infty } ^ { \infty } | \psi ( t ) | ^ { 2 } d t = 1 }.\end{array}
\end{equation}
Through scaling and shifting, a family of wavelet functions is then obtained by
\begin{equation}
	\psi _ { a , b } ( t ) = \frac { 1 } { \sqrt { | a | } } \psi ( \frac { t - b } { a } ),
\end{equation}
where $a$ represents the scaling factor and $b$ denotes the translation factor. The continuous wavelet transform of a signal $f(t)$ is performed by computing the inner product between the signal and this family of wavelet functions, namely
\begin{equation}
	W _ { f } ( a , b ) = \langle f , \psi _ { a , b } \rangle = \frac { 1 } { \sqrt { | a | } } \int _ { - \infty } ^ { \infty } f ( t ) \psi ^ { * } ( \frac { t - b } { a } ) d t,
\end{equation}
where the asterisk $(*)$ denotes the complex conjugate. In practical computation, the continuous scaling and shifting factors $(a,b)$ lead to substantial redundant calculations. To address this problem, discrete scaling and shifting factors are employed to generate a series of discrete functions. Typically, the scaling factor varies in powers of 2, namely
\begin{equation}
	a = 2 ^ { j } , \quad j \in Z,
\end{equation}
where $j$ denotes the scale series or number of layers. Larger values of $j$ correspond to wider low-frequency wavelets, while smaller values yield narrower high-frequency wavelets. The translation factor $b$ is set to $2^j \times k$ with integer $k$, ensuring that
coarser scales use larger sampling steps and finer scales use smaller steps. This choice aligns with the principle of multi-resolution analysis and yields a non-redundant, complete computation. The resulting discrete wavelet functions are
\begin{equation}
	\psi _ { j , k } ( t ) = \frac { 1 } { \sqrt { 2 ^ { j } } } \psi ( \frac { t - k \cdot 2 ^ { j } } { 2 ^ { j } } ) = 2 ^ { - j / 2 } \psi ( 2 ^ { - j } t - k ).
\end{equation}
The discrete wavelet series expansion of a signal $f(t)$ then takes the form
\begin{equation}
	f ( t ) = \sum _ { j = - \infty } ^ { \infty } \sum _ { k = - \infty } ^ { \infty } d _ { j , k } \psi _ { j , k } ( t ),
\end{equation}
where the wavelet coefficients are given by
\begin{equation}
	d _ { j , k } = \langle f , \psi _ { j , k } \rangle = \int _ { - \infty } ^ { \infty } f ( t ) \psi _ { j , k } ( t ) d t.
\end{equation}
In practice, the lowest-frequency components of a signal are bounded, and an infinite number of basis functions is computationally infeasible. These issues are addressed within the multi-resolution analysis framework, which decomposes the function space into a nested sequence of approximation subspaces
\begin{equation}
	V _ {  j - 1} = V _ {j}  \oplus W _ { j } .
\end{equation}
Here, $V _ {j} $ denotes the approximation subspace, and $W _ { j }$ represents the detail subspace. The approximation subspace is spanned by the scaling functions $\phi (t)$ and the detail subspace by the wavelet functions $\psi (t)$. Any signal $f(t)$ can be expressed up to a coarsest scale $J$ as
\begin{equation}
	f ( t ) = \sum _ { k } c _ { J , k } \phi _ { J , k } ( t ) + \sum _ { j = J } ^ { \infty } \sum _ { k } d _ { j , k } \psi _ { j , k } ( t ),
	\label{wavelet}
\end{equation}
where the scaling functions $\phi (t)$ are constructed analogously, namely
\begin{equation}
	\phi _ { j , k } ( t ) = 2 ^ { -j / 2 } \phi ( 2 ^ { -j } t - k ).
	\label{waveletmom}
\end{equation}

\section{\label{sec:PIKAN}Physics-informed Kolmogorov-Arnold network (PIKAN)}
Kolmogorov-Arnold network (KAN) is inspired by the Kolmogorov-Arnold representation theorem \cite{schmidt2021kolmogorov}, which states that any multivariate continuous function defined on a compact domain can be represented by finite compositions of continuous univariate functions. Specifically, for a smooth function $\bm{f}: [0,1]^n \rightarrow \mathbb{R}$, the theorem gives
\begin{equation}
	\bm{f}(\bm{x})=
	\bm{f}(x_1,\cdots,x_n)
	=
	\sum_{q=1}^{2n+1}
	\Phi_q
	\left(
	\sum_{p=1}^{n}
	\phi_{q,p}(x_p)
	\right),
\end{equation}
where $\phi_{q,p}: [0,1] \mapsto \mathbb{R}$ and $\Phi_q:\mathbb{R} \mapsto  \mathbb{R}$ are continuous univariate functions. Unlike conventional multilayer perceptrons, which place nonlinear activation functions on neurons, KAN introduces learnable univariate functions on network edges. This provides a flexible representation in which both the nonlinear transformation and the corresponding functional basis are learned from data.

For the $l$-th KAN layer, let $\bm{x}_{l}\in\mathbb{R}^{N_l}$ and $\bm{x}_{l+1}\in\mathbb{R}^{N_{l+1}}$ denote the input and output feature vectors, respectively. The layer mapping can be written as
\begin{equation}
	x_{l+1,j}
	=
	\sum_{i=1}^{N_l}
	\phi_{j,i}^{(l)}(x_{l,i}),
	\qquad
	j=1,2,\cdots,N_{l+1},
\end{equation}
or equivalently,
\begin{equation}
	\bm{x}_{l+1}
	=
	\Phi_l(\bm{x}_l)
	=
	\begin{bmatrix}
		\phi_{1,1}^{(l)}(\cdot) & \phi_{1,2}^{(l)}(\cdot) & \cdots & \phi_{1,N_l}^{(l)}(\cdot) \\
		\phi_{2,1}^{(l)}(\cdot) & \phi_{2,2}^{(l)}(\cdot) & \cdots & \phi_{2,N_l}^{(l)}(\cdot) \\
		\vdots & \vdots & \ddots & \vdots \\
		\phi_{N_{l+1},1}^{(l)}(\cdot) & \phi_{N_{l+1},2}^{(l)}(\cdot) & \cdots & \phi_{N_{l+1},N_l}^{(l)}(\cdot)
	\end{bmatrix}
	\bm{x}_l ,
\end{equation}
where $\Phi_l$ denotes the collection of learnable edge functions in the $l$-th KAN layer. Each univariate function $\phi_{j,i}^{(l)}(\cdot)$ is parameterized by a residual activation component and a spline component as
\begin{equation}
	\phi_{j,i}^{(l)}(z)
	=
	w_{b,j i}^{(l)} b(z)
	+
	w_{s,j i}^{(l)}
	{\rm spline}_{j i}^{(l)}(z),
\end{equation}
where $z$ is a scalar input, $w_{b,j i}^{(l)}$ and $w_{s,j i}^{(l)}$ are trainable scaling coefficients, and the spline term is expressed as a linear combination of B-spline basis functions,
\begin{equation}
	{\rm spline}_{j i}^{(l)}(z)
	=
	\sum_{m=1}^{M}
	c_{j i m}^{(l)} B_m(z).
\end{equation}
Here, $B_m(z)$ denotes the $m$-th B-spline basis function, and $c_{j i m}^{(l)}$ are learnable spline coefficients. The residual activation is chosen as the SiLU function,
\begin{equation}
	b(z)
	=
	{\rm SiLU}(z)
	=
	\frac{z}{1+\exp(-z)}.
\end{equation}

In the physics-informed Kolmogorov-Arnold network (PIKAN) considered in this work, the KAN serves as the surrogate model for the solution field,
\begin{equation}
	\bm{u}_{\theta}^{\rm KAN}(\bm{x},t)
	=
	{\rm KAN}_{\theta}(\bm{x},t),
\end{equation}
where $\theta$ collects all trainable parameters, including the spline coefficients and scaling factors in the KAN layers. The physical constraints are imposed in the same manner as in the standard PINN formulation described in Sec.~\ref{sec:level3}. Specifically, the initial-condition, boundary-condition, and governing-equation residual losses are combined into the composite objective function defined in Eq.~\ref{16}. The PDE residual is evaluated by substituting $\bm{u}_{\theta}^{\rm KAN}(\bm{x},t)$ into the governing equation, namely
\begin{equation}
	\mathcal{L}_{\rm pde}
	\left(
	\bm{u}_{\theta}^{\rm KAN};
	\bm{x}_{\rm pde},t_{\rm pde}
	\right)
	=
	\frac{\partial \bm{u}_{\theta}^{\rm KAN}}
	{\partial t}
	(\bm{x}_{\rm pde},t_{\rm pde})
	-
	R
	\left(
	\bm{u}_{\theta}^{\rm KAN};
	\bm{x}_{\rm pde},t_{\rm pde}
	\right),
\end{equation}
with the required derivatives computed by automatic differentiation. Therefore, PIKAN differs from the standard PINN mainly in the choice of the neural architecture: the fully connected MLP backbone in PINN is replaced by a KAN backbone, while the physics-informed training objective remains consistent with the formulation in Sec.~\ref{sec:level3}.

\begin{table}[h!]
	\centering
	\footnotesize
	\setlength{\tabcolsep}{4pt}
	\caption{Relative $L^2$ errors for the shallow water equations at different spatial locations.}
	\label{tab:shallow_water_ablation_errors}
	\renewcommand{\arraystretch}{1.1}
	\resizebox{\textwidth}{!}{%
	\begin{tabular}{lcccccccc}
		\toprule
		Model / Position & $x=$0 & 200 & 400 & 600 & 800 & 1000 & 1200 & Average \\
		\midrule
		$h$ (PIWF)    & $8.499 \times 10^{-5}$ & $5.009 \times 10^{-4}$ & $1.494 \times 10^{-3}$ & $8.833 \times 10^{-4}$ & $1.603 \times 10^{-3}$ & $1.947 \times 10^{-3}$ & $2.357 \times 10^{-3}$ & $8.779 \times 10^{-4}$ \\
		$h$ (PIWF-nonF) & $8.834 \times 10^{-1}$ & $8.093 \times 10^{-1}$ & $7.136 \times 10^{-1}$ & $6.415 \times 10^{-1}$ & $5.925 \times 10^{-1}$ & $5.632 \times 10^{-1}$ & $5.469 \times 10^{-1}$ & $7.745 \times 10^{-1}$ \\
		$h$ (PIWF-nonW) & $1.154 \times 10^{0}$ & $3.147 \times 10^{0}$ & $5.365 \times 10^{0}$ & $7.739 \times 10^{0}$ & $1.096 \times 10^{1}$ & $1.311 \times 10^{1}$ & $1.548 \times 10^{1}$ & $8.213 \times 10^{0}$ \\
		$h$ (PIWF-nonL) & $1.982 \times 10^{-3}$ & $4.513 \times 10^{-3}$ & $5.254 \times 10^{-3}$ & $8.217 \times 10^{-3}$ & $9.662 \times 10^{-3}$ & $1.101 \times 10^{-2}$ & $1.580 \times 10^{-2}$ & $4.759 \times 10^{-3}$ \\
		$h$ (PINN)    & $4.981 \times 10^{-5}$ & $4.068 \times 10^{-2}$ & $4.517 \times 10^{-2}$ & $4.829 \times 10^{-2}$ & $5.347 \times 10^{-2}$ & $5.989 \times 10^{-2}$ & $6.344 \times 10^{-2}$ & $3.685 \times 10^{-2}$ \\
		$h$ (PINN-tri) & $3.611 \times 10^{-4}$ & $6.945 \times 10^{-3}$ & $1.128 \times 10^{-2}$ & $1.295 \times 10^{-2}$ & $1.331 \times 10^{-2}$ & $1.026 \times 10^{-2}$ & $8.380 \times 10^{-3}$ & $6.040 \times 10^{-3}$ \\
		\midrule
		$hu$ (PIWF)   & $8.652 \times 10^{-5}$ & $4.557 \times 10^{-4}$ & $1.415 \times 10^{-3}$ & $6.686 \times 10^{-4}$ & $1.373 \times 10^{-3}$ & $1.721 \times 10^{-3}$ & $2.155 \times 10^{-3}$ & $8.742 \times 10^{-4}$ \\
		$hu$ (PIWF-nonF) & $3.668 \times 10^{-1}$ & $3.493 \times 10^{-1}$ & $3.772 \times 10^{-1}$ & $4.299 \times 10^{-1}$ & $5.068 \times 10^{-1}$ & $6.026 \times 10^{-1}$ & $7.116 \times 10^{-1}$ & $6.692 \times 10^{-1}$ \\
		$hu$ (PIWF-nonW) & $8.518 \times 10^{-1}$ & $9.952 \times 10^{-1}$ & $5.743 \times 10^{0}$ & $8.773 \times 10^{0}$ & $7.893 \times 10^{0}$ & $1.556 \times 10^{1}$ & $2.875 \times 10^{1}$ & $8.746 \times 10^{0}$ \\
		$hu$ (PIWF-nonL) & $1.987 \times 10^{-3}$ & $3.855 \times 10^{-3}$ & $4.019 \times 10^{-3}$ & $6.108 \times 10^{-3}$ & $7.076 \times 10^{-3}$ & $7.916 \times 10^{-3}$ & $1.210 \times 10^{-2}$ & $4.347 \times 10^{-3}$ \\
		$hu$ (PINN)   & $1.953 \times 10^{-6}$ & $5.378 \times 10^{-3}$ & $1.074 \times 10^{-2}$ & $1.512 \times 10^{-2}$ & $2.185 \times 10^{-2}$ & $3.021 \times 10^{-2}$ & $3.599 \times 10^{-2}$ & $9.899 \times 10^{-3}$ \\
		$hu$ (PINN-tri) & $3.746 \times 10^{-4}$ & $7.128 \times 10^{-3}$ & $1.157 \times 10^{-2}$ & $1.327 \times 10^{-2}$ & $1.356 \times 10^{-2}$ & $1.038 \times 10^{-2}$ & $8.380 \times 10^{-3}$ & $6.093 \times 10^{-3}$ \\
		\bottomrule
	\end{tabular}%
	}
\end{table}

\section{\label{sec:ablation}Ablation experiment}
We conduct an ablation experiment and a parameter sensitivity analysis of the PIWF model in this section. Three PIWF-based control models are considered: PIWF without the wavelet layer (PIWF-nonW), PIWF without the Fourier-basis layer (PIWF-nonF), and PIWF without the MLP layer (PIWF-nonL). Since each layer of PIWF contains three parallel feature extraction sublayers, the model appears to have three times as many neurons as a conventional neural network. Although the neurons among the three feature extraction sublayers of our PIWF model do not connect to each other, we still include a PINN model with three times the width (PINN-tri) as a control for rigor. Table \ref{tab:shallow_water_ablation_errors} reports the relative errors for the shallow water equation. The frozen models all exhibit an increase in error to varying degrees compared to the original model, which demonstrates the necessity of introducing the wavelet layer, the Fourier-basis layer, and the MLP layer. In this example, the PIWF-nonW model shows the most significant error increase, followed by PIWF-nonF, and then PIWF-nonL. For the shallow water equation with a discontinuous solution, the wavelet layer plays a key role in capturing local features, therefore, PIWF-nonW exhibits a larger error than PIWF-nonL. The results also indicate that the absence of either Fourier or wavelet feature extraction causes the model to overemphasize either global or local features, leading to training imbalance and a substantial increase in the final error. The PIWF-nonL model shows a relatively small error increase since the Fourier and wavelet feature extractions capture most of the important features, leaving the residual feature extraction to account for a minor portion. Notably, the three-times-width PINN (PINN-tri) shows a clear advantage over the standard PINN. Despite the absence of information exchange among the different feature extraction sublayers, PIWF still maintains a significant advantage over PINN-tri, which demonstrates that the superiority of PIWF does not arise simply from stacking more neurons but from a more efficient training architecture compared to conventional neural networks.

\begin{table}[h!]
	\centering
	\footnotesize
	\setlength{\tabcolsep}{4pt}
	\caption{Relative $L^2$ errors for the two-dimensional cylinder wake flow at $\rm Re=100$.}
	\label{tab:cylinder_errors}
	\renewcommand{\arraystretch}{1.1}
	\resizebox{\textwidth}{!}{%
	\begin{tabular}{lcccccccc}
		\toprule
		Model / Time & 1 & 2 & 3 & 4 & 5 & 6 & 7 & Average \\
		\midrule
		$u$ (PIWF)  & $9.769 \times 10^{-4}$ & $6.584 \times 10^{-4}$ & $4.716 \times 10^{-4}$ & $3.589 \times 10^{-4}$ & $2.830 \times 10^{-4}$ & $2.373 \times 10^{-4}$ & $2.106 \times 10^{-4}$ & $4.567 \times 10^{-4}$ \\
		$u$ (PIWF-nonF) & $2.571 \times 10^{-3}$ & $1.637 \times 10^{-3}$ & $1.084 \times 10^{-3}$ & $7.605 \times 10^{-4}$ & $5.776 \times 10^{-4}$ & $4.943 \times 10^{-4}$ & $4.961 \times 10^{-4}$ & $1.089 \times 10^{-3}$ \\
		$u$ (PIWF-nonW) & $3.857 \times 10^{-4}$ & $3.434 \times 10^{-4}$ & $3.266 \times 10^{-4}$ & $3.269 \times 10^{-4}$ & $3.373 \times 10^{-4}$ & $3.595 \times 10^{-4}$ & $3.758 \times 10^{-4}$ & $3.507 \times 10^{-4}$ \\
		$u$ (PIWF-nonL) & $6.551 \times 10^{-4}$ & $5.629 \times 10^{-4}$ & $5.499 \times 10^{-4}$ & $5.276 \times 10^{-4}$ & $5.023 \times 10^{-4}$ & $4.929 \times 10^{-4}$ & $4.870 \times 10^{-4}$ & $5.397 \times 10^{-4}$ \\
		$u$ (PINN)  & $6.074 \times 10^{-3}$ & $4.178 \times 10^{-3}$ & $2.956 \times 10^{-3}$ & $2.212 \times 10^{-3}$ & $1.693 \times 10^{-3}$ & $1.360 \times 10^{-3}$ & $1.172 \times 10^{-3}$ & $2.806 \times 10^{-3}$ \\
		$u$ (PINN-tri) & $4.176 \times 10^{-3}$ & $2.832 \times 10^{-3}$ & $1.972 \times 10^{-3}$ & $1.412 \times 10^{-3}$ & $1.029 \times 10^{-3}$ & $7.812 \times 10^{-4}$ & $6.237 \times 10^{-4}$ & $1.832 \times 10^{-3}$ \\
		\midrule
		$v$ (PIWF)  & $2.212 \times 10^{-3}$ & $1.593 \times 10^{-3}$ & $1.141 \times 10^{-3}$ & $9.184 \times 10^{-4}$ & $7.809 \times 10^{-4}$ & $7.054 \times 10^{-4}$ & $6.929 \times 10^{-4}$ & $1.149 \times 10^{-3}$ \\
		$v$ (PIWF-nonF) & $5.658 \times 10^{-3}$ & $4.019 \times 10^{-3}$ & $2.854 \times 10^{-3}$ & $2.220 \times 10^{-3}$ & $1.812 \times 10^{-3}$ & $1.559 \times 10^{-3}$ & $1.373 \times 10^{-3}$ & $2.785 \times 10^{-3}$ \\
		$v$ (PIWF-nonW) & $1.115 \times 10^{-3}$ & $1.032 \times 10^{-3}$ & $9.195 \times 10^{-4}$ & $9.040 \times 10^{-4}$ & $9.141 \times 10^{-4}$ & $9.055 \times 10^{-4}$ & $8.754 \times 10^{-4}$ & $9.522 \times 10^{-4}$ \\
		$v$ (PIWF-nonL) & $1.349 \times 10^{-3}$ & $1.122 \times 10^{-3}$ & $1.090 \times 10^{-3}$ & $1.042 \times 10^{-3}$ & $9.585 \times 10^{-4}$ & $9.019 \times 10^{-4}$ & $9.082 \times 10^{-4}$ & $1.053 \times 10^{-3}$ \\
		$v$ (PINN)  & $1.212 \times 10^{-2}$ & $9.613 \times 10^{-3}$ & $7.710 \times 10^{-3}$ & $6.396 \times 10^{-3}$ & $5.447 \times 10^{-3}$ & $4.790 \times 10^{-3}$ & $4.354 \times 10^{-3}$ & $7.204 \times 10^{-3}$ \\
		$v$ (PINN-tri) & $7.568 \times 10^{-3}$ & $5.785 \times 10^{-3}$ & $4.323 \times 10^{-3}$ & $3.419 \times 10^{-3}$ & $2.817 \times 10^{-3}$ & $2.485 \times 10^{-3}$ & $2.213 \times 10^{-3}$ & $4.087 \times 10^{-3}$ \\
		\midrule
		$p$ (PIWF)  & $1.802 \times 10^{-3}$ & $9.964 \times 10^{-4}$ & $1.090 \times 10^{-3}$ & $1.013 \times 10^{-3}$ & $8.990 \times 10^{-4}$ & $8.393 \times 10^{-4}$ & $8.027 \times 10^{-4}$ & $1.063 \times 10^{-3}$ \\
		$p$ (PIWF-nonF) & $3.570 \times 10^{-3}$ & $2.260 \times 10^{-3}$ & $2.088 \times 10^{-3}$ & $1.869 \times 10^{-3}$ & $2.125 \times 10^{-3}$ & $2.076 \times 10^{-3}$ & $1.784 \times 10^{-3}$ & $2.253 \times 10^{-3}$ \\
		$p$ (PIWF-nonW) & $9.982 \times 10^{-4}$ & $1.114 \times 10^{-3}$ & $1.442 \times 10^{-3}$ & $1.059 \times 10^{-3}$ & $1.050 \times 10^{-3}$ & $1.141 \times 10^{-3}$ & $9.876 \times 10^{-4}$ & $1.113 \times 10^{-3}$ \\
		$p$ (PIWF-nonL) & $2.335 \times 10^{-3}$ & $1.421 \times 10^{-3}$ & $1.434 \times 10^{-3}$ & $1.509 \times 10^{-3}$ & $1.504 \times 10^{-3}$ & $1.307 \times 10^{-3}$ & $1.276 \times 10^{-3}$ & $1.541 \times 10^{-3}$ \\
		$p$ (PINN)  & $1.254 \times 10^{-2}$ & $9.112 \times 10^{-3}$ & $9.144 \times 10^{-3}$ & $6.225 \times 10^{-3}$ & $5.983 \times 10^{-3}$ & $5.134 \times 10^{-3}$ & $4.141 \times 10^{-3}$ & $7.469 \times 10^{-3}$ \\
		$p$ (PINN-tri) & $6.494 \times 10^{-3}$ & $4.796 \times 10^{-3}$ & $5.878 \times 10^{-3}$ & $3.434 \times 10^{-3}$ & $3.024 \times 10^{-3}$ & $2.224 \times 10^{-3}$ & $2.678 \times 10^{-3}$ & $4.075 \times 10^{-3}$ \\
		\bottomrule
	\end{tabular}%
	}
\end{table}

\begin{table}[h!]
	\centering
	\caption{Training cost comparison in GPU hours for different models and problems.}
	\label{tab:training_time}
	\resizebox{\textwidth}{!}{%
	\begin{tabular}{lcccc}
		\toprule
		Model / Case & Burgers ($\nu=0.01/\pi$) & Burgers ($\nu=0.001$) & Shallow water equation & Cylinder flow wake \\
		\midrule
		PINN  & $0.45$  & $0.45$  & $3.20$  & $3.52$ \\
		PIKAN & $5.81$  & $5.53$  & $4.26$  & $48.06$ \\
		PIWF  & $5.30$  & $5.30$  & $5.63$  & $2.88$ \\
		\bottomrule
	\end{tabular}%
	}
\end{table}

Table \ref{tab:cylinder_errors} reports the relative errors for the cylinder flow wake. All frozen models again exhibit error increases to varying degrees, confirming the necessity of each feature layer. Compared to the shallow water equation case that contains a discontinuous solution, the cylinder flow wake favors the capture of global features. Consequently, in this example, PIWF-nonF shows a more pronounced error increase than PIWF-nonW, which cross-validates the necessity of having both the Fourier-basis layer and the wavelet layer in different examples. The PINN-tri model still does not achieve a significant performance improvement over the standard PINN, and PIWF retains a substantial advantage. Finally, we report the training time for each model in each example in Table \ref{tab:training_time}. The PIWF model does not exhibit a significantly longer training time than the other models in most cases.

\section*{References}
\bibliographystyle{aipnum4-2}
\bibliography{ref.bib}
\end{document}